\RequirePackage{amsmath}
\documentclass[final]{amsart}
\usepackage[utf8x]{inputenc}
\usepackage{graphicx}
\usepackage{hyperref}
\usepackage{amsmath,amssymb}
\usepackage{textcomp}
\usepackage[total={6.5in,8in}]{geometry}
\usepackage{color}
\usepackage{todonotes}

\title{Hybrid reaction-diffusion and clock-and-wavefront model for the arrest of oscillations in the somitogenesis segmentation clock}

\author{Jes\'us Pantoja-Hern\'andez\textsuperscript{1} \and V\'ictor F. Bre\~na-Medina\textsuperscript{2} \and Mois\'es Santill\'an\textsuperscript{1}}
\address{\textsuperscript{1}Centro de Investigaci\'on y de Estudios Avanzados del IPN, Unidad Monterrey, V\'ia del Conocimiento 201, Parque PIIT, 66628 Apodaca NL, M\'exico.}
\address{\textsuperscript{2}Department of Mathematics, ITAM, R\'io Hondo 1, Ciudad de M\'exico 01080, M\'exico.}

\date{\today}

\begin{document}
\maketitle
\pagestyle{myheadings}
\thispagestyle{plain}
\markboth{J.~Pantoja-Hern\'andez, V.~Bre\~na--Medina \and M.~Santill\'an}
{Hybrid reaction-diffusion and clock-and-wavefront model for the arrest of oscillations in the somitogenesis segmentation clock} 
	

	\begin{abstract}
	The clock and wavefront paradigm is arguably the most widely accepted model for explaining the embryonic process of somitogenesis. According to this model, somitogenesis is based upon the interaction between a genetic oscillator, known as segmentation clock, and a differentiation wavefront, which provides the positional information indicating where each pair of somites is formed. Shortly after the clock and wavefront paradigm was introduced, Meinhardt presented a conceptually different mathematical model for morphogenesis in general, and somitogenesis in particular. Recently, Cotterell \emph{et al.} rediscovered an equivalent model by systematically enumerating and studying small networks performing segmentation. Cotterell \emph{et al.} called it a progressive oscillatory reaction-diffusion (PORD) model. In the Meinhardt-PORD model, somitogenesis is driven by short-range interactions and the posterior movement of the front is a local, emergent phenomenon, which is not controlled by global positional information. With this model, it is possible to explain some experimental observations that are incompatible with the clock and wavefront model. However the Meinhardt-PORD model has some important disadvantages of its own. Namely, it is quite sensitive to fluctuations and depends on very specific initial conditions (which are not biologically realistic). In this work, we propose an equivalent Meinhardt-PORD model, and then amend it to couple it with a wavefront consisting of a receding morphogen gradient. By doing so, we get a hybrid model between the Meinhardt-PORD and the clock-and-wavefront ones, which overcomes most of the deficiencies of the two originating models.
	\end{abstract}
	
	\maketitle
	
	\textbf{
Somitogenesis, the process by which somites are formed, is an essential developmental stage in many vertebrates. This process occurs with a strikingly regular periodicity, that is preserved among embryos of a single species. The clock and wavefront paradigm is arguably the most widely accepted model for explaining somitogenesis. However, it is incapable of explaining some experimental facts, like the appearance of somites in the absence of an external wavefront (i.e. a receding morphogen gradient). Shortly after the clock and wavefront paradigm was introduced, Meinhardt presented a conceptually different mathematical model for morphogenesis in general, and somitogenesis in particular. Recently, Cotterell \emph{et al.} rediscovered an equivalent model by systematically enumerating and studying small networks performing segmentation, and called it a progressive oscillatory reaction-diffusion (PORD) model. The Meinhardt-PORD model tackles some of the deficiencies of the clock and wavefront models, but it has some serious issues of its own. In the present work, we introduce an equivalent Meinhardt-PORD model, and then amend it to couple it with a receding morphogen gradient. By doing so, we get a hybrid model that incorporates characteristics of the Meinhardt-PORD and clock-and-wavefront models. We show that this hybrid model undergoes a bifurcation, from a stable to an unstable limit cycle, as the value of the parameter accounting for a background regulatory input (associated to the receding morphogen gradient) decreases. This bifurcation allows the model to explain why somites can form in the absence of an external wavefront, reassesses the role of the receding morphogen gradient as a conductor for somitogenesis, and makes the model behavior robust to random fluctuations, as  well as independent from specific initial conditions (the latter, are two of the weak points of the Meinhardt-PORD model). We argue that this findings provide convincing evidence that reaction-diffusion and positional information (receding morphogen gradient) mechanisms could work together in somitogenesis.	
}

\section{Introduction}
\label{intro}

Somitogenesis, the process by which somites are formed, is an essential developmental stage in many species. Somites are bilaterally paired blocks of mesoderm cells that form along the anterior-posterior axis of the developing embryo in segmented animals \cite{Maroto2012}. In vertebrates, somites give rise to skeletal muscle, cartilage, tendons, endothelial cells, and dermis. Somites form with a strikingly regular periodicity, that is preserved among embryos of a single species. From this, and other reasons, scientists have been attracted to somitogenesis for decades. One of the earliest conceptual attempts to explain this regularity is the so-called clock and wavefront model, originally proposed by Cooke and Zeeman~\cite{Cooke1976}. According to this model, somitogenesis occurs due to the interaction between: (i) autonomous oscillations of a network of genes and gene products, which causes presomitic-mesoderm cells to oscillate between a permissive and a non-permissive state, in a consistently (clock-like) timed fashion; and (ii) an external wavefront (also known as determination front) of signaling that slowly progresses in an anterior-to-posterior direction. As the wavefront comes in contact with cells in the permissive state, they undergo a mesenchymal-epithelial transition, forming a somite boundary, and resetting the process for the next somite.
	
The clock and wavefront model gained relevance when the expression of several genes under the Notch, Wnt, and FGF pathways in mice and chicken, as well as genes like \textit{her/hes} in all vertebrates \cite{Schorter2012}, was discovered to oscillate cyclically with the same period as that of somite formation \cite{Palmeirim1997, Pourquie2001, Gibb2010, Pourquie2011}. This, together with the existence of morphogens, whose concentrations vary along the presomitic mesoderm in characteristic patterns that travel at constant velocity in an anterior to posterior direction (receding morphogen gradients), seemed to confirm the general assessments of the clock and wavefront model \cite{Gibb2010, Pourquie2011, Dubrulle2001}. As a matter of fact, this model has been so successful because it agrees with numerous experimental observations, and because much progress has been done in elucidating the segmentation-clock clockwork, and the way it interacts with morphogen gradients. Despite its success, some experimental observations have been reported that are incompatible with the clock and wavefront paradigm. For instance, somites can form in the absence of morphogen gradients, albeit in a disorderly fashion \cite{Naiche2011,Dias2014}.
	
Shortly after the clock and wavefront paradigm was introduced, Meinhardt presented a conceptually different mathematical model for morphogenesis in general, and somitogenesis in particular \cite{Meinhardt1982}. Contrarily to the clock and wavefront model, in the Meinhardt model, somitogenesis is driven by short-range interactions. Hence, the posterior movement of the front is a local, emergent phenomenon, that is not controlled by global positional information. See also \cite{Francois2018} for a thorough review of the Meinhardt model. Interestingly, a dynamically equivalent model was recently rediscovered by Cotterell \emph{et al.}~\cite{Cotterell2015} by systematically enumerating and studying small networks performing segmentation. Cotterell \emph{et al.} called their model a progressive oscillatory reaction-diffusion (PORD) system. Here and thereafter, we will refer to it as the Meinhardt-PORD model. As discussed by Meinhardt \cite{Meinhardt1982}, Fran\c{c}ois \cite{Francois2018}, and Cotterell \emph{et al.} \cite{Cotterell2015}, the he Meinhardt-PORD model is compatible with some important features of somitogenesis that former reaction-diffusion models were unable to explain. Furthermore, the Meinhardt-PORD model makes predictions regarding FGF-inhibition and tissue-cutting experiments that are more consistent with experimental observations than those of clock and wavefront models.
	
	\begin{figure}[t!]
		\centering
		\includegraphics[width=2in]{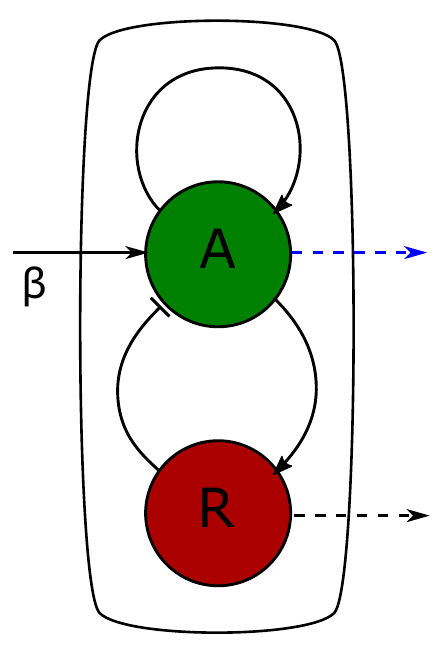}
		\caption{Schematic representation of the gene regulatory network introduced by
			\cite{Cotterell2015} and studied in the present work. This network consists of
			two genes: an activator \textit{A} and a repressor \textit{R}. Solid lines
			ending in arrowheads (hammerheads) denote positive (negative) regulation.
			$\beta$ represents external regulation of gene \textit{A} via a 
			receding morphogen gradient. Dashed lines correspond to passive diffusion into the extracellular
			medium. The diffusion process depicted by the blue arrow was not
			originally included in the Cotterell {\em et al.}~model, but is 
			included in the modified model version studied in this work.}
		\label{Fig01}
	\end{figure}
	
At the core of the Meinhardt-PORD model there is a gene regulatory motif schematically depicted in Fig.~\ref{Fig01}. In this network, a hypothetical gene codes for an activator protein~(A), which enhances its own expression, as well as that of another gene that codes for a hypothetical repressor protein~(R). In turn, the repressor protein down-regulates the gene coding for the activator, and is able to diffuse into the extracellular medium and affect neighboring cells. The dynamic behavior of the gene network in Fig. \ref{Fig01} is consistent with most of the experimental observations regarding somitogenesis. In particular, with the correct parameter values, it can generate sustained oscillations, and when repressor diffusion is included, it gives rise to a stationary pattern of gene expression like that observed in somitogenesis, even in the absence of an external wavefront.
	
In spite of the Meihardt-PORD model being hypothetical, in the sense that an oscillatory gene network with the topology illustrated in Fig. \ref{Fig01} has not been identified so far, we find it interesting for the following reasons: (i)~as far as we have observed, it accounts for some important features of somitogenesis that traditional clock and wavefront models fail to explain; (ii)~a gene network with a similar architecture as that of Fig. \ref{Fig01}---albeit with a different dynamic behavior---has been invoked to explain oscillation arrest in somitogenesis \cite{Santillan2008, Zavala2012}; and finally, (iii)~the architecture of the gene network in Fig. \ref{Fig01} is an ubiquitous motif in the intricate transcription-factor regulatory network of the genes under the Wnt, Notch and FGF pathways \cite{Gibb2010, Zavala2012}.
	
In opposition to its multiple virtues, the Meihardt-PORD model has some important issues of its own. For instance, it is quite sensitive to initial-condition fluctuations. This implies that a disordered pattern of gene expression arises when small-amplitude random perturbations of initial conditions are present. This behavior resembles what happens with PSM cells that are not under the influence of a receding morphogen gradient, but contrasts with the observed robustness of somite formation under normal conditions. Another issue of the Meinhardt-PORD model is that it requires very specific initial conditions (which are not biologically realistic) to give rise to a segmentation pattern. 

We speculate from the above discussion that, although an external wavefront is not strictly necessary for the Meihardt-PORD model to generate a gene expression pattern consistent with somitogenesis, the interaction with a receding morphogen gradient is essential to explain the observed robustness of this phenomenon and to eliminate its dependence on specific initial conditions. In other words, we argue that a hybrid model that accounts for the diffusion of the regulatory proteins (as in the Meihardt-PORD model), as well as for an external wavefront (in the form of a receding morphogen gradient) that interacts with the oscillatory gene network (as in the clock and wavefront paradigm), may circumvent the deficiencies of the original models. The present work is aimed at testing this hypothesis, and discussing the corresponding biological implications.
	
The manuscript is organized as follows. In section \ref{model} we introduce an equivalent Meinhardt-PORD model, and expand it to couple it with a receding morphogen gradient. In section \ref{param} we define the parameter set where the dynamics of the non-diffusion system give place to a limit cycle family. In section \ref{numer} we explain the numerical methods used, as well as the initial and boundary conditions. In section \ref{res} we present and discuss the present work results. Finally, we discuss the relevance of the obtained results, as well as the limitations of the model in section \ref{conclu}.
	
	\section{Mathematical Model}
	\label{model}
	
Consider the gene network schematically represented in Fig. \ref{Fig01}. Assume that the half life of mRNA molecules corresponding to genes \textit{A} and \textit{R} is much shorter than that of the corresponding proteins. Then, a quasi-stationary approximation can be made for the equations governing mRNA dynamics, which yields the following reaction-diffusion system for the concentration of proteins $A$ and $R$, under the assumption that both proteins diffuse across cell membranes and into the extracellular medium:
	\begin{subequations}\label{eq012}
		\begin{flalign}
		& \frac{\partial A}{\partial t} = \alpha_A P_A (A, R) - \mu_A A + D_A \nabla^2 A\,,
		\label{eq01} \\
		& \frac{\partial R}{\partial t} = \alpha_R P_R (A) - \mu_R R + D_R \nabla^2 R\,,
		\label{eq02}
		\end{flalign}
	\end{subequations}
where $\alpha_A$ and $\alpha_R$ are the maximum possible rate of activator and repressor production; $P_A$ and $P_B$ are the probabilities that the promoters of genes $A$ and $R$ are active; degradation rate constants of each protein concentration are denoted by $\mu_A$ and $\mu_R$, respectively; and $D_A$ and $D_R$ are the corresponding diffusion coefficients.

It is important to emphasize at this point that $D_A = 0$ in the Meinhardt-PORD model. However, we consider for the sake of generality that both the repressor and the activator can diffuse. Later on, we show that introducing activator diffusion is necessary to efficiently couple the system with a receding morphogen gradient and get a clock-and-wavefront behavior.
	
To take into account the roles of the activator and the repressor, $P_A(A, R)$ must be a monotonic increasing (decreasing) function of $A$ ($R$), while $P_R(A)$ ought to be a monotonic increasing function of $A$. 

Meinhardt~\cite{Meinhardt1982} assumed that $P_A(A, R) = \displaystyle \rho_1 A^2/R$ and $P_R(A) = \rho_2 A^2$, where $\rho_{1,2}>0$.
These expressions are troublesome from a biochemical perspective because $P_A(A, R)$ diverges as $A \to \infty$ while, in real life, gene expression rates saturate for large activator concentrations. On the other hand, both $P_A(A, R)$ and $P_R(A)$ diverge as $R \to 0$ as can be seen from noticing that: (i)~as $P_A(A,R)$ quadratically increases unbounded as the represor positively decays to naught, which hence (ii)~promotes an increasing growth of the activator as $\partial A/\partial t>0$ to consequently get an unbounded $P_R(A)$ growth. However, this is not the way gene expression behaves in the absence of repressors.

On their own, Cotterell {\em et al.}~\cite{Cotterell2015} proposed the following functions:
	\begin{subequations}\label{eq034}
		\begin{gather}
		P_A(A, R) = \displaystyle \Phi \left(
		\frac{l_1 A - l_2 R + \beta}{1 + l_1 A - l_2 R + \beta}
		\right), \label{eq03} \\[3mm]
		P_R(A) = \displaystyle \frac{l_3 A}{1 + l_3 A}, \label{eq04}
		\end{gather}
	\end{subequations}
where $l_1$, $l_2$, and $l_3$ define the strengths of regulatory interactions between \textit{A} and \textit{R}, and $\beta$ is the background regulatory input of \textit{A}. To prevent negative values, Cotterell {\em et al.}~introduced the function $\Phi(x) = x.H(x)$, where $H(x)$ is the standard Heaviside function ($H(x) = 1$ for $x \geq 0$ and $H(x) = 0$ for $x < 0$). Even though the function $P_A(A, R)$ defined in (\ref{eq03}) fulfills the requirement of being a monotonic decreasing function of $R$ and a monotonic increasing function of $A$, it shows features that are biologically challenging:
\begin{itemize}
\item There is neither biological nor biochemical motivation for the introduction of function~$\Phi$.

\item Function $\Phi(x)$ is non smooth at the origin. This feature is unusual in biologically inspired mathematical models, and may cause unexpected complications while studying the dynamical system; for further discussion of oscillatory behaviors on non-smooth dynamical systems, see for instance~\cite{han01,castillo01}.

\item Since the term $l_2 R$ appears with negative sign in the denominator of the argument of function $\Phi$, in the right hand side of Eq. (\ref{eq03}), function $P_A(A, R)$ may be divergent for certain values of $A$ and $R$.
\end{itemize}
In this work, we propose a slightly novel approach by introducing functions $P_A(A, R)$ and $P_R(A)$ that are consistent with the assumptions that the activator and the repressor compete for the same binding site in the promoter region of gene $A$, and that both the activator and the repressor interact with their corresponding binding sites in the promoter regions of genes $A$ and $R$ in a cooperative fashion \cite{Santillan2008b}. In so doing, we have
	\begin{subequations}\label{eq056}
		\begin{gather}
		P_A(A,R) = \displaystyle \frac{\beta + (A/K_1)^{n_1}}{1 + (A/K_1)^{n_2} +
			(R/K_2)^{n_2}}\,, \label{eq05} \\[3mm]
		P_R(A)  = \displaystyle \frac{(A/K_3)^{n_3}}{1 + (A/K_3)^{n_3}}\,,
		\label{eq06}
		\end{gather}
	\end{subequations}
where $K_1$ denotes the half saturation constant for the binding reaction between the activator and the promoter of gene $A$; the Hill coefficient that accounts for a cooperative interaction between the activator and gene $A$ promoter is given by $n_1$; as in the original model, $\beta$ is the background regulatory input of $A$; the half saturation constant and Hill coefficient of the interaction between the repressor and the promoter of gene $A$ are represented by $K_2$ and $n_2$, respectively; and $K_3$ and $n_3$, equivalently, are the half saturation constant and Hill coefficient for the interaction between the activator and the promoter of gene $R$. 
	
Notice that \eqref{eq012} along with \eqref{eq056} constitute a reaction-diffusion system for the gene expression network depicted in Fig.~\ref{Fig01}, which accounts for spatio-temporal interactions between these two protein concentrations.
	
Upon re-scaling position and time as $x' = x / L$, $y' = y / L$, $z' = z / L$, and $t' = t \mu_A$ and substituting the following dimensionless variables and parameters 
	\begin{gather*}
	a  =  \dfrac{A \mu_A}{\alpha_A},  \qquad  d_a  =  \dfrac{D_A}{ \mu_A L^2}, 
	\qquad k_1  =  \dfrac{K_1 \mu_A }{ \alpha_1},  \qquad k_2  =  \dfrac{K_2 \mu_R
	}{\alpha_3},  \\
	r  =  \dfrac{R \mu_r }{ \alpha_r}, \qquad  d_r  = \dfrac{ D_R }{ \mu_A L^2},
	\qquad k_3  =  \dfrac{K_3 \mu_R }{ \alpha_R},  \qquad  \mu  = \dfrac{ \mu_R 
	}{\mu_A}, 
	\end{gather*}
where $L$ is a characteristic length on system \eqref{eq012}-\eqref{eq056}, we obtain the reaction-diffusion system in a dimensionless form 
	\begin{subequations}\label{eq078910}
		\begin{gather}
		\frac{\partial a}{\partial t'} =  P_a(a, r) - a + d_a \nabla'^2 a, \label{eq07} \\
		\frac{\partial r}{\partial t'}  =  \mu\left (P_r(a) - r + d_r \nabla'^2 r\right),
		\label{eq08}
		\end{gather}
		where $\nabla'$ is the Laplacian with respect to $(x', y', z')$, and
		\begin{gather}
		P_a(a, r)  =  \displaystyle \frac{\beta + (a / k_1)^{n_1}}{1 + (a /
			k_1)^{n_1} + (r / k_2)^{n_2}}, \label{eq09} \\[2mm]
		P_r(a)  =  \frac{(a / k_3)^{n_3}}{1 + (a / k_3)^{n_3}}. \label{eq10}
		\end{gather}
	\end{subequations}
To ease notation, we suppress symbol $(\cdot)^\prime$ from this point onward along the paper.
	
\section{Parameter estimation}
\label{param}
	
Since the genes here modeled are hypothetical, it is impossible to estimate the model parameter values from experimental data. Instead, we performed a bifurcation analysis of the system with no diffusion ($d_a = d_r = 0$), employing a continuation method implemented in \texttt{XPPAUT} \cite{Ermentrout1987}. The results of this analysis are presented in Fig.~\ref{FigA01} of Appendix \ref{app:bif}. From this analysis, we determined the parameter intervals (see Table \ref{Tab01}) for which the system shows sustained oscillations in the absence of diffusion, which is crucial to set the periodicity of somitogenesis. In our simulations, we consider parameter values in the middle of the intervals reported in Table \ref{Tab01}. Upon following \cite{Cotterell2015}, we fixed $n_3 = \mu = 1$, and assumed that $n = n_1 = n_2$, to then leave the diffusion coefficients as the only free parameters in the model. 
	
	\begin{table}[h] 
		\centering
		\begin{tabular}{|c|} \hline
			$k_1  \in [0.029, 0.057]$ \\ \hline
			$k_2  \in [0.008, 0.017]$ \\ \hline 
			$k_3  \in [1.658, 3.642]$ \\ \hline
			$n  \in  [2.411, 5.025]$ \\  \hline
			$\beta  \in  (0, 2.25]$ \\   \hline
		\end{tabular} 
		
		\caption{Parameter intervals for which the system with no diffusion shows
			sustained stable oscillatory behavior.}
		\label{Tab01}
	\end{table}
	
\section{Numerical methods}
\label{numer}
	
Under the supposition that the presomitic mesoderm (PSM) can be regarded as one-dimensional, we considered a single spacial dimension, $x$, with boundaries at $x=0$ and $x=1$. These boundaries set an observation window in the PSM, where $x=0$ corresponds to the posterior extreme. To numerically solve system~\eqref{eq078910}, we implemented a standard finite-difference three-point stencil and Euler's algorithm in \texttt{Julia}; homogeneous Neumann boundary conditions were included:
	\begin{gather}\label{eqbn}
	\left. \dfrac{\partial a}{\partial x}\right|_{(0, t)} = 
	\left. \dfrac{\partial a}{\partial x}\right|_{(1, t)} = 
	\left. \dfrac{\partial r}{\partial x}\right|_{(0, t)} =
	\left. \dfrac{\partial r}{\partial x}\right|_{(1, t)} = 0 ,
	\end{gather}
		
We also performed stochastic simulations in which additive white noise was added to the system. For these simulations we substituted equations \eqref{eq07} and
	\eqref{eq08} by
	\begin{subequations}\label{eq1112}
		\begin{gather}
		\frac{\partial a}{dt} = P_a(a, r) - a + d_a \nabla^2 a + \text{CV}_{\text{noise}} \, a^* \,\frac{dW}{dt}, \label{eq11} \\
		\frac{\partial r}{dt}= \mu (P_r(a) + r - d_r \nabla^2 r) + \text{CV}_{\text{noise}} \, r^* \,\frac{dW}{dt}, \label{eq12}
		\end{gather}
	\end{subequations}
where $a^*$ and $r^*$ respectively denote the  values of  $a$ and $r$ at the unstable homogeneous steady state, $\text{CV}_{\text{noise}}$ is the coefficient of variation of the added white noise, and $dW/dt$ is a normally-distributed white noise term with mean zero and variance one. To solve this system of stochastic partial differential equations we employed the Euler-Maruyama method, implemented in \texttt{Julia}.
	
We employed different initial conditions along the present work and they are specified along each simulation.

\section{Results}
\label{res}
	
\begin{figure}[ht!]
		\centering
		\includegraphics[width=2in]{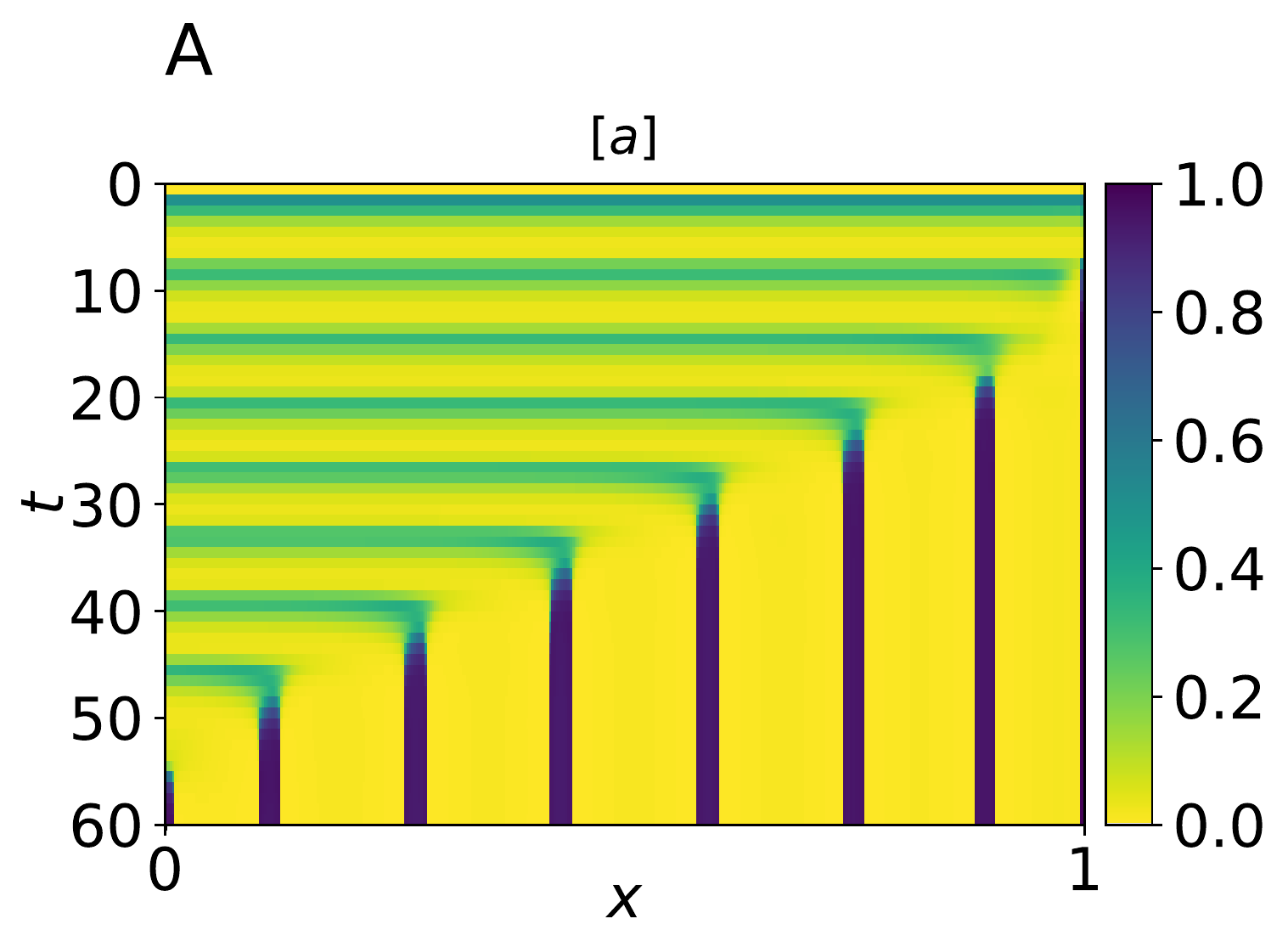}
		\includegraphics[width=2in]{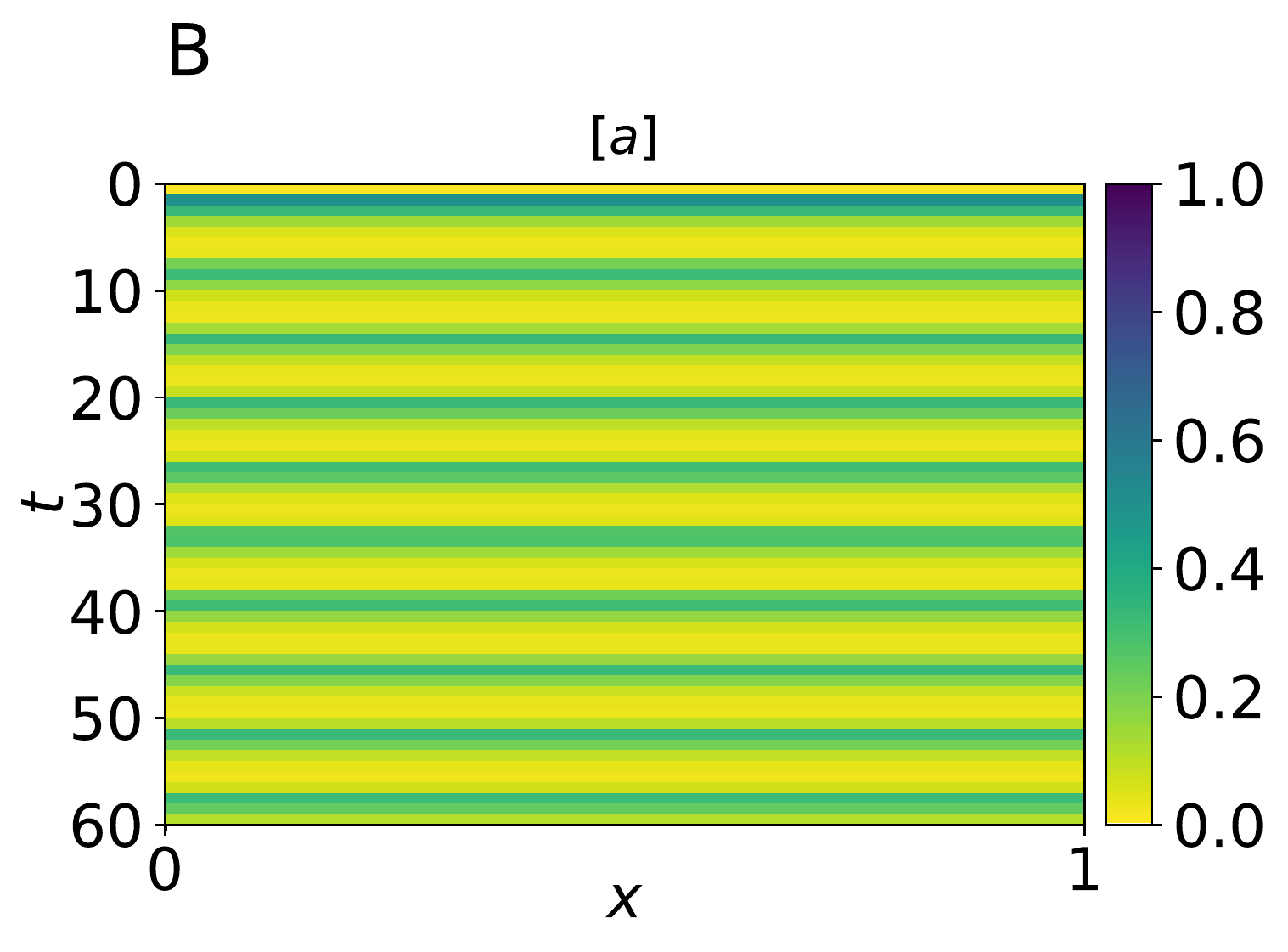}
		\includegraphics[width=2in]{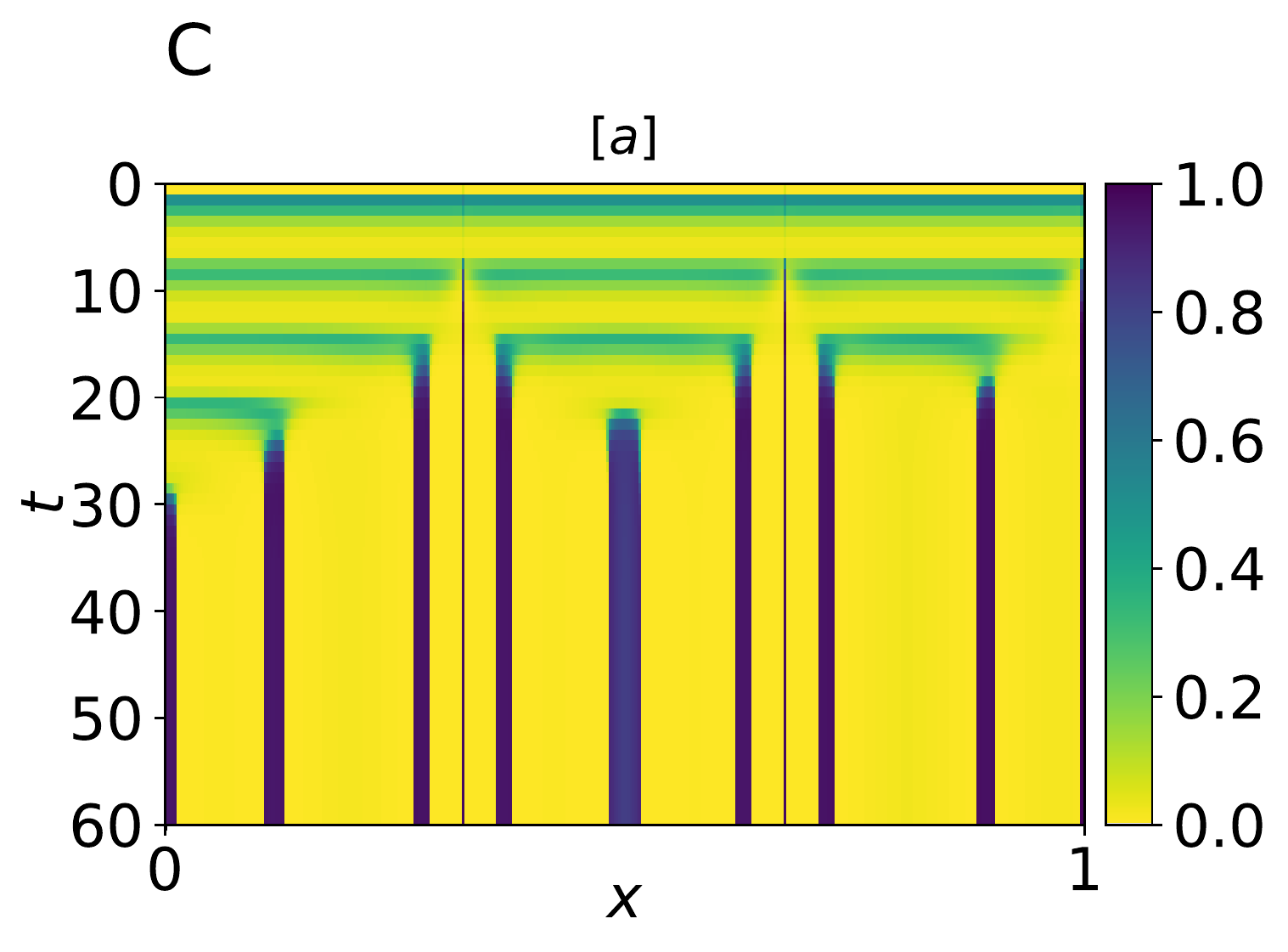}
		\caption{Spatio-temporal evolution of $a$ from system~\eqref{eq078910}, with boundary condition as in~\eqref{eqbn}, for different initial conditions: (A)$r(x, 0)=0$ for all $x\in[0,1]$, $a(x, 0)=0$ for $0\leq x<1$, and $a(1, t) = 0.05$; (B) $r(x, 0) = a(x, 0)=0$ for all $x\in[0,1]$; (C) $r(x, 0)=0$ and $a(x, 0)=0$ for all $x\in[0,1]$ and $a(x',0)=0.05$ at $x'=0.325, 0.675, 0.9975$. Parameter values were set as follows: $k_1=0.05$, $k_2=0.01$, $k_3=2$, $n=3$, $\beta=0.5$, $d_a=0$, and $d_r=2.5\times10^{-3}$. Animations with the same data sets used to plot the heat-maps in this figure can be found in the reservoir \url{https://github.com/JesusPantoja/Reaction-Diffusion_Movies/}.}
		\label{Fig02}
	\end{figure}

We started by reproducing the results in \cite{Cotterell2015} to test the equivalence of the Meinhardt-PORD model and the present one, in the absence of activator diffusion. To this end, we set $d_a=0$, $d_r = 2.5\times10^{-3}$, $\beta = 0.5$, and numerically solved the model equations as described in Section~\ref{numer}, with the following initial conditions:
	\begin{gather}
	a(x, 0) = \left\{\begin{array}{cl}
	0.05 & \text{for } x = 1, \\
	0 & \text{for } 0 \leq x < 1,
	\end{array} \right. 
	\quad
	r(x, 0) = 0, \; \text{for } 0 \leq x \leq 1\,.
	\end{gather}
That is, the system is assumed to be initially homogeneous, except for a perturbation at the anterior extreme of the observation window. The simulation results for the activator concentration are shown in Fig.~\ref{Fig02}A. To better visualize the temporal evolution of the animations in this work, we created animations with the data sets, here plotted as heat maps, and put them in the public reservoir \url{https://github.com/JesusPantoja/Reaction-Diffusion_Movies/}. Observe that an oscillatory behavior gradually gives rise to a steady pattern, consisting of alternated high and low activator concentration regions, in agreement with the reported behavior of the Meinhardt \cite{Meinhardt1982} and PORD \cite{Cotterell2015} models. 
	
Further notice in Fig.~\ref{Fig02}A that, once the system has reached a stationary pattern, the high-activator-concentration stripes are much shorter than those corresponding to low activator concentration. This behavior is different from that of the Meindhardt-PORD model, which renders alternated equally-sized stripes. Meindhart interpreted them as corresponding to the anterior and posterior phenotypes (AP) of a somite. Meinhardt also pointed out that this pattern was not sufficient to explain somitogenesis, as another periodic stripe (S) was needed, at least, to determine somite boundaries via the pattern: SAPSAPSAPS, and suggested that a more complex gene network was required to reproduce this pattern. In our model, we propose that a high activator level is the signal that triggers the mesenchymal-epithelial transition, and so the high-activator-concentration stripes can be interpreted as corresponding to the boundaries between adjacent somites (S stripes). Hence, somites occupy the low-activator-level regions in the final stationary pattern. As in the Meindhart model, ours does not explain all segmentation features. In particular, it does not account for the differentiation between the anterior and posterior somite phenotypes, and a more complex network would be required to do so. Interestingly, Dunty et al. \cite{Dunty2008} have demonstrated that the Wnt3a/$\beta$-catenin pathway is permissive but not instructive for oscillating clock genes and that it controls the anterior-posterior positioning of boundary formation in the presomitic mesoderm (PSM). 

By looking closely at the dynamics pictured in Fig.~\ref{Fig02}A, we can observe that the pattern formation dynamics start at the initial perturbation position, and propagate with constant speed. The initial perturbation is essential for the appearance of the pattern. If the system is initially homogeneous, the oscillatory behavior continues indefinitely---see Fig.~\ref{Fig02}B. On the other hand, when more than one initial perturbations are present, each one of them originates a pattern-formation wave, and when two such waves collide, they cancel out---see Fig.~\ref{Fig02}C. These results suggest that the transition from an oscillatory behavior to a steady pattern of gene expression is due to a diffusion driven instability interacting with a limit cycle. To verify this, we investigated the spatial stability of the dynamical system in~\eqref{eq078910} and~\eqref{eqbn} in Appendix \ref{app:bif}. We were able to confirm that the limit cycle is unstable with the current parameter values set; see Fig.~\ref{FigB02}, panel (c), at $\kappa^2=0$, where $\kappa$ denotes the Fourier mode, also known as wave mode.
	
The results described in the previous paragraph, which are qualitatively equivalent to those previously reported by Meinhardt~\cite{Meinhardt1982} and Cotterell \emph{et al.}~\cite{Cotterell2015}, are consistent with the reported experimental observations that somites can form in the absence of an external wavefront (recall that somite formation in the absence of an external wavefront is incompatible with the clock and wavefront paradigm). As a matter of fact, they may explain why somites form almost simultaneously and irregularly; i.e. any initial perturbation in the mesoderm tissue would rapidly originate a somite boundary, and the emerging pattern formation wave would almost immediately collide with neighboring waves. On the other hand, the results in Fig.~\ref{Fig02}C disagree with the observed robustness of somitogenesis to many different kinds of perturbations on both the mesoderm tissue and the differentiation wavefront. Moreover, Meinhardt-PORD models demand very specific initial conditions to produce ordered segmentation patterns, but this is not biologically realistic. 

There are several reports that confirm the importance of the differentiation wavefront \cite{Naiche2011, Sawada2001}. In particular, it has been reported in chick embryos that ``FGF8 is sufficient to maintain the caudal identity of presomitic mesoderm cells and that down-regulation of FGF8 signalling at the level of the determination front is required to enable cells to proceed further with the segmentation process'' \cite{Dubrulle2001}. We speculate from this that, although a receding morphogen gradient is not strictly necessary in Meinhardt-PORD models to generate a segmentation pattern, somitogenesis robustness can be achieved by coupling the model with a receding morphogen gradient, and that this will also eliminate the stringent dependence on a specific initial condition. In this regard, Meinhardt \cite{Meinhardt1982} demonstrated that coupling its model with a morphogen gradient can bias reaction-diffusion-based patterning. However, he considered a static gradient instead of the differentiation wavefront which is observed in vertebrates.
	
	\begin{figure}[t!]
		\centering
		\includegraphics[width=2in]{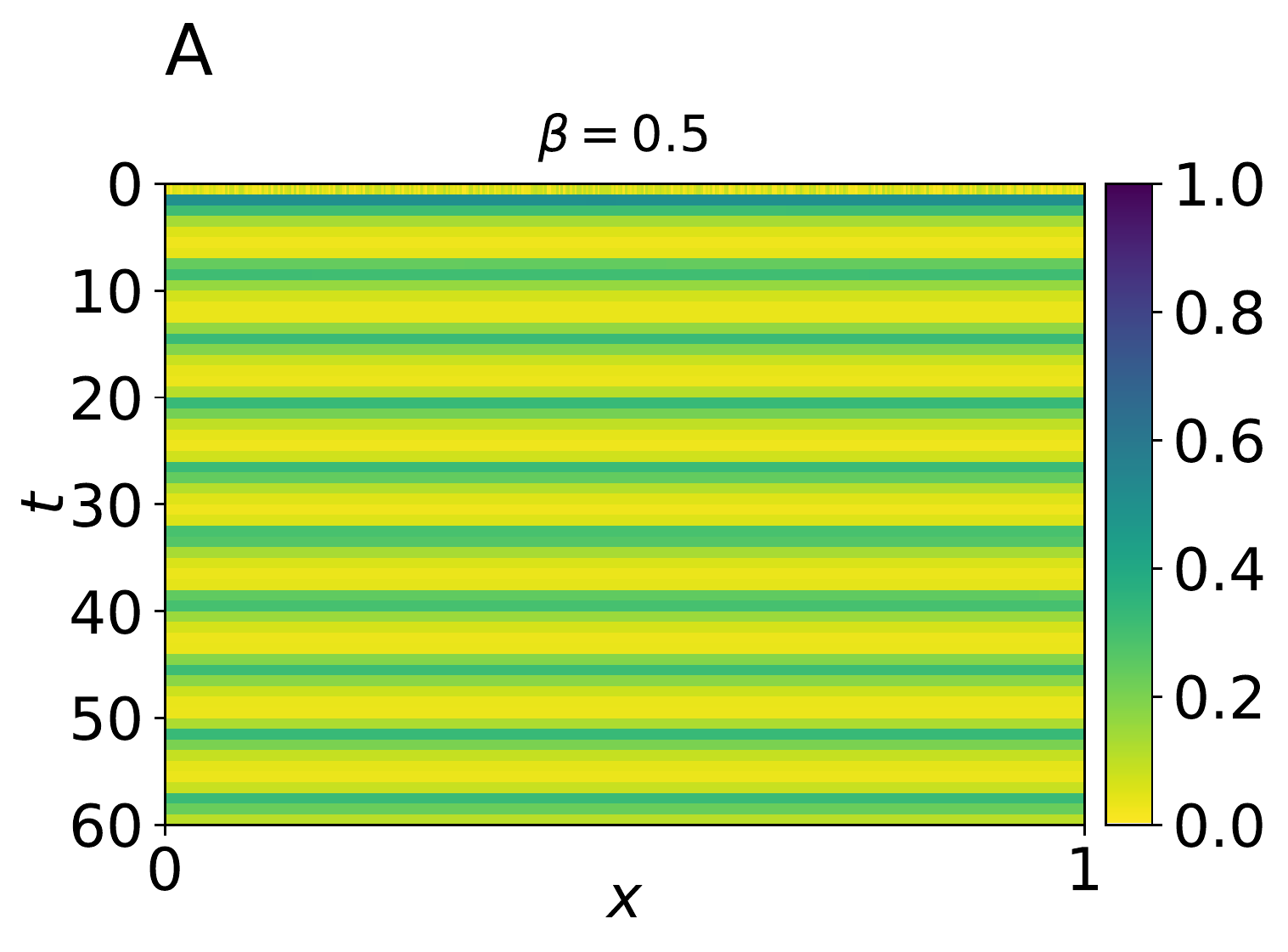}
		\includegraphics[width=2in]{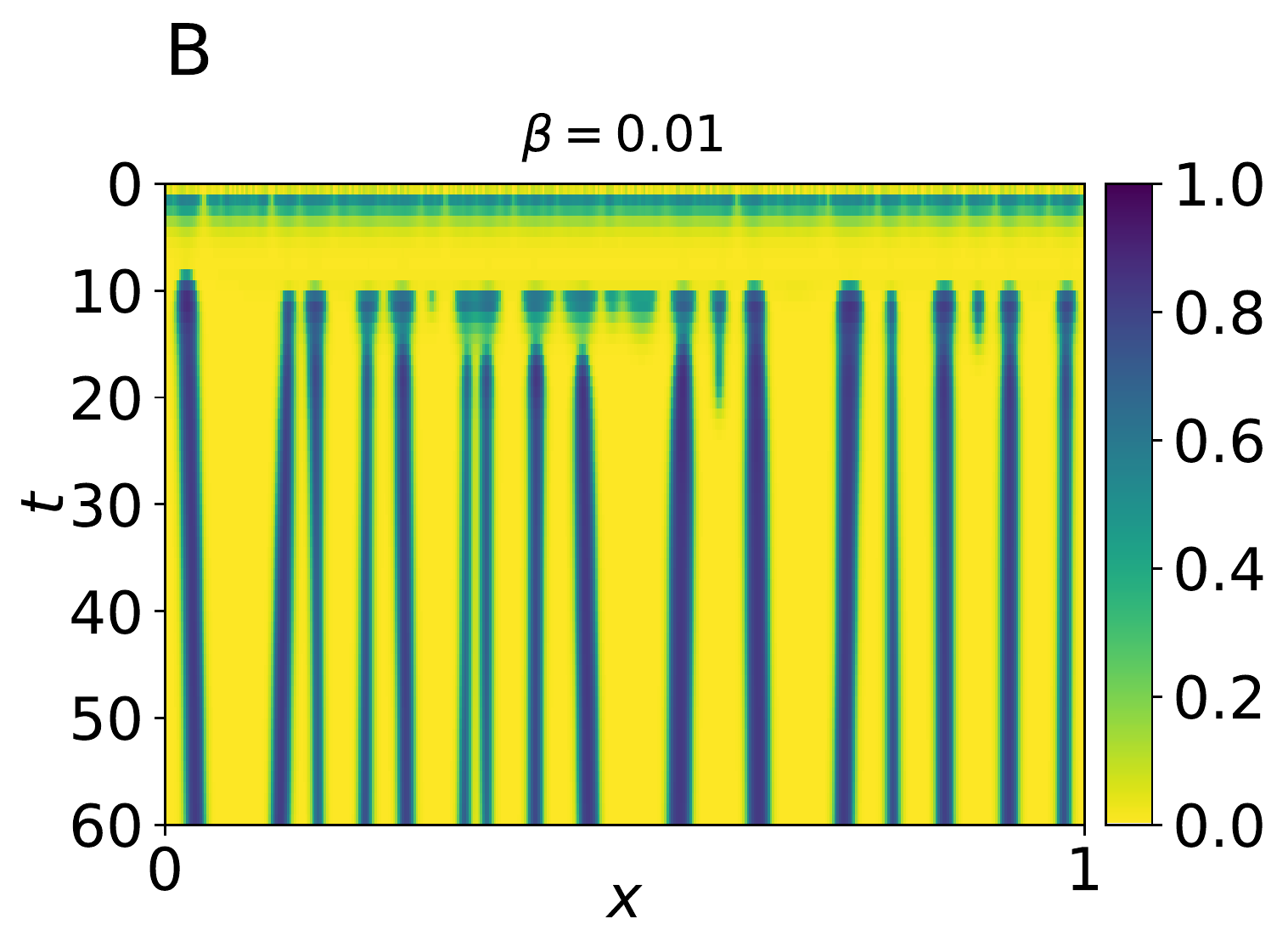}
		\caption{Spatio-temporal evolution of $a$ from system~\eqref{eq078910}, with boundary condition as in~\eqref{eqbn}, for: (A) spatially stable with $\beta=0.5$; (B) spatially unstable with $\beta=0.01$. In both cases, the initial conditions for $a$ were selected from random uniform distributions in the interval $[0, 0.1]$, whereas we set $r(x, 0) = 0$ for all $x\in[0,1]$. Parameter values were set as follows: $k_1=0.05$, $k_2=0.01$, $k_3=2$, $n=3$, $d_a = 5\times10^{-5}$ and $d_r=2.5\times10^{-3}$. Animations with the same data sets used to plot the heat-maps in this figure can be found in the reservoir \url{https://github.com/JesusPantoja/Reaction-Diffusion_Movies/}.}
		\label{Fig03}
	\end{figure}

It is commonly accepted that the differentiation wavefront is originated by morphogens (like FGF8 and Wnt3a in chicken and mice) which are produced in the embryo tail bud and diffuse to the rest of the PSM. In consequence, the morphogen concentration decreases in the posterior to anterior direction. Furthermore, as the embryo grows, the tail bud recedes leaving PSM cells behind. Hence, the spatial morphogen distribution moves in the anterior to posterior direction, like a wavefront, as time passes. Taking this into account, together with the fact that, following Cotterell \emph{et al.} \cite{Cotterell2015}, parameter $\beta$ accounts for the gene network interaction with the external morphogen, we speculate that the model in \eqref{eq078910} might be coupled with a $\beta$ wavefront, to yield a clock-and-wavefront behavior, provided that the system has a spatially stable limit cycle for large $\beta$ values, and turns unstable below a given $\beta$ threshold. In this way, large morphogen levels would maintain an oscillatory gene expression despite perturbations, thus preventing somite formation close to the tail bud. Furthermore, once $\beta$ goes below the threshold at which the limit cycle becomes unstable, any local inhomogeneity would lead to the formation of a somite at the PSM position where the threshold is reached.

	\begin{figure}[t!]
		\centering
		\includegraphics[width=2in]{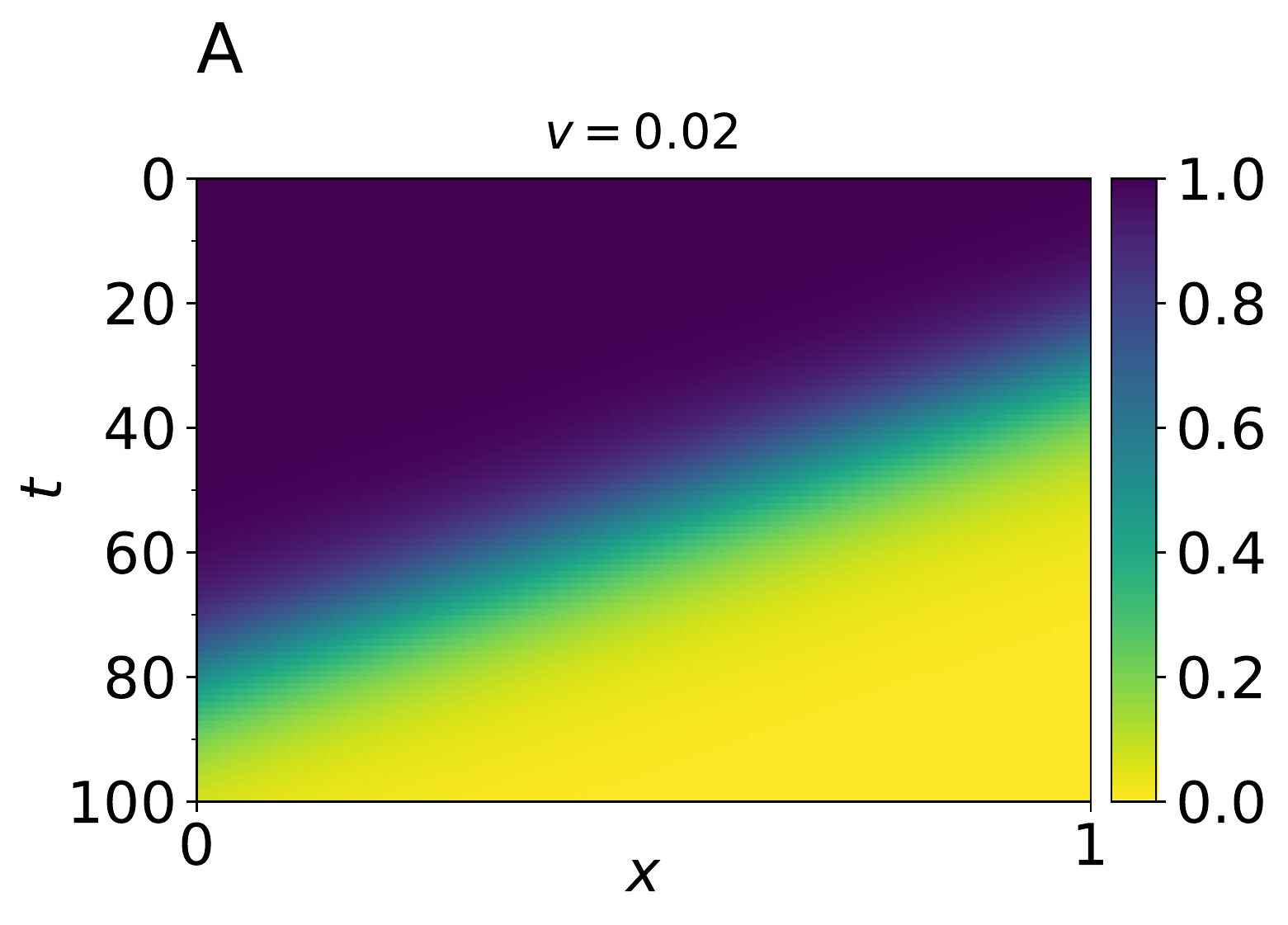}
		\includegraphics[width=2in]{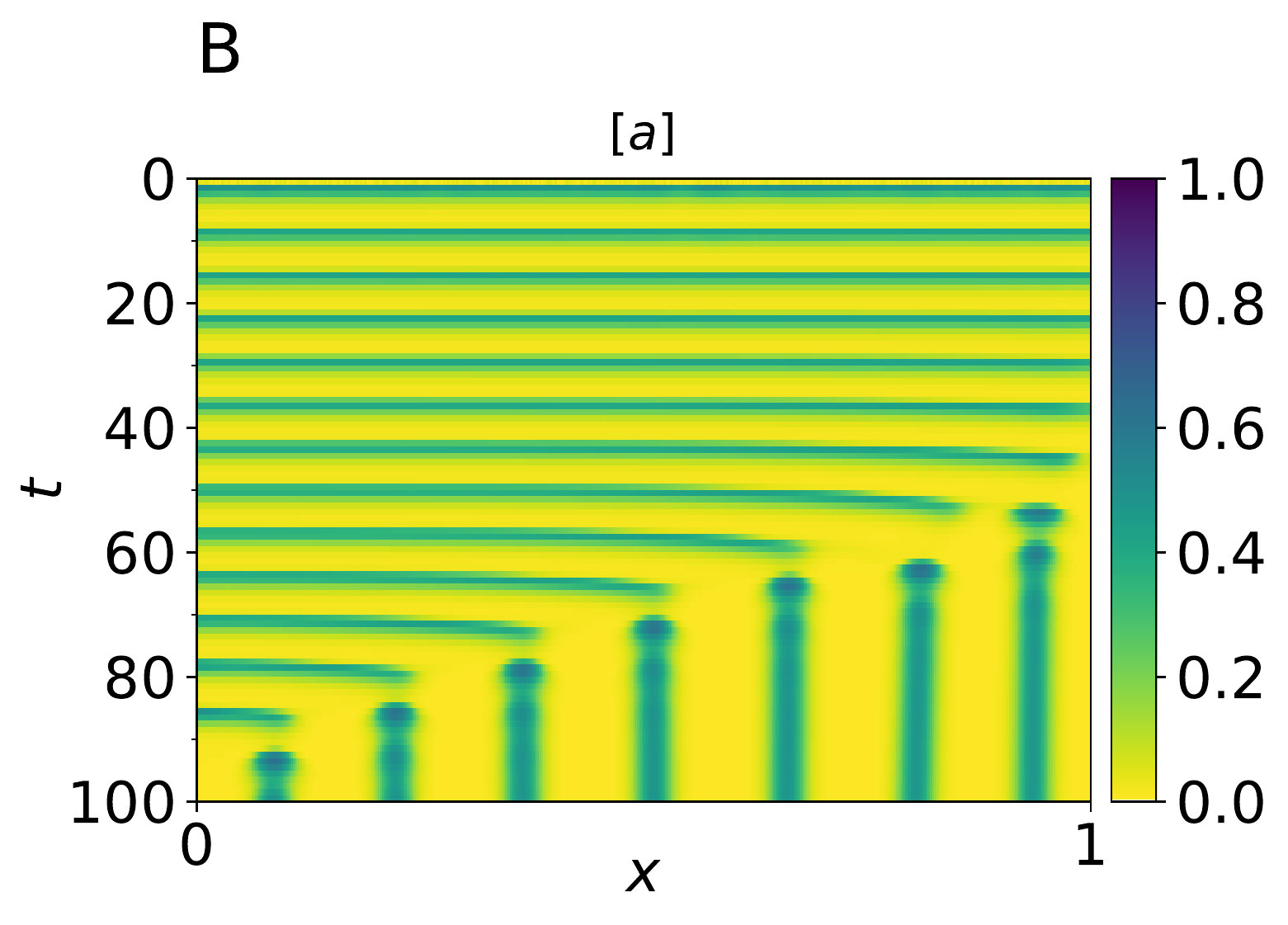} \\
		\includegraphics[width=2in]{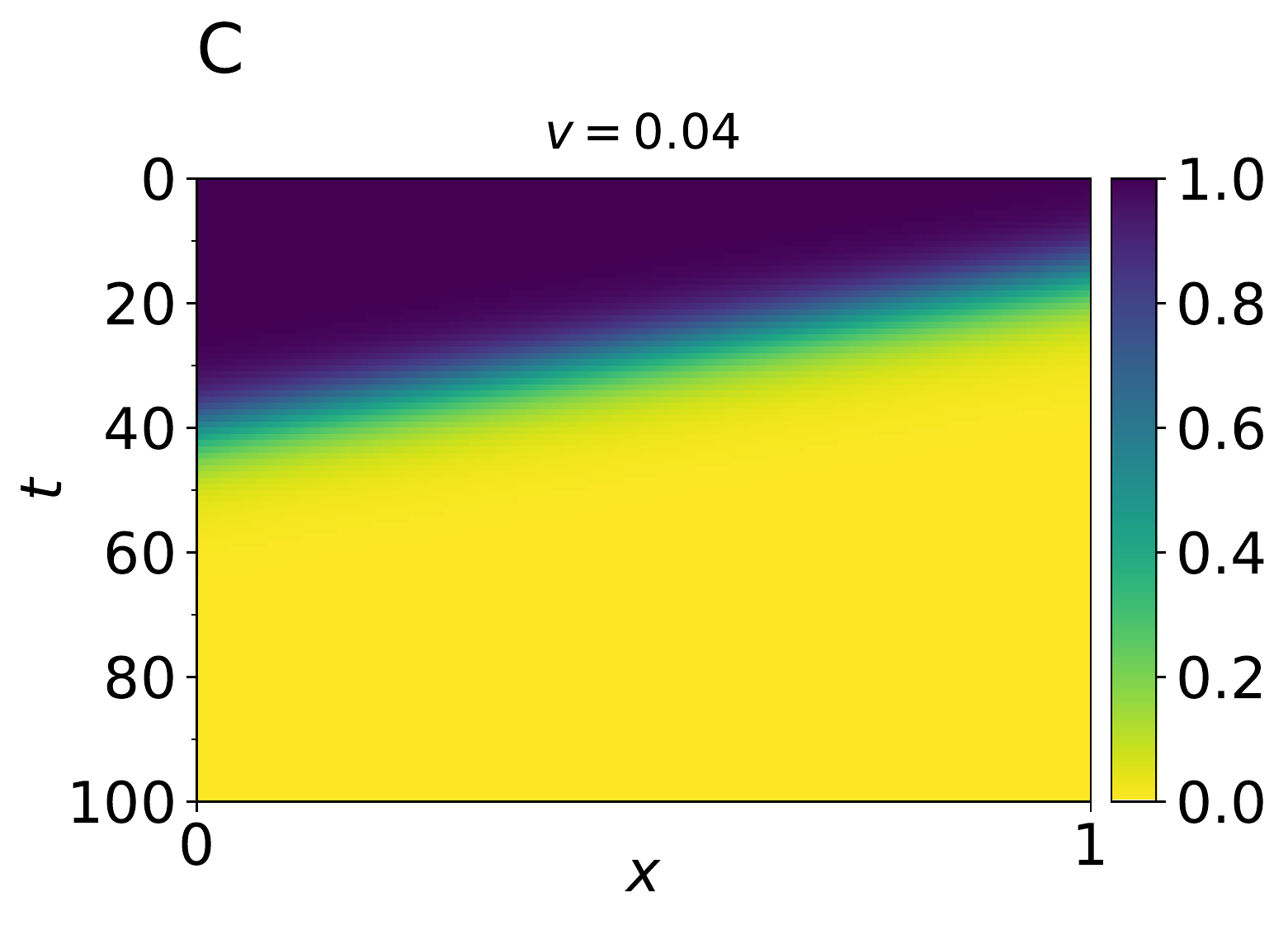}
		\includegraphics[width=2in]{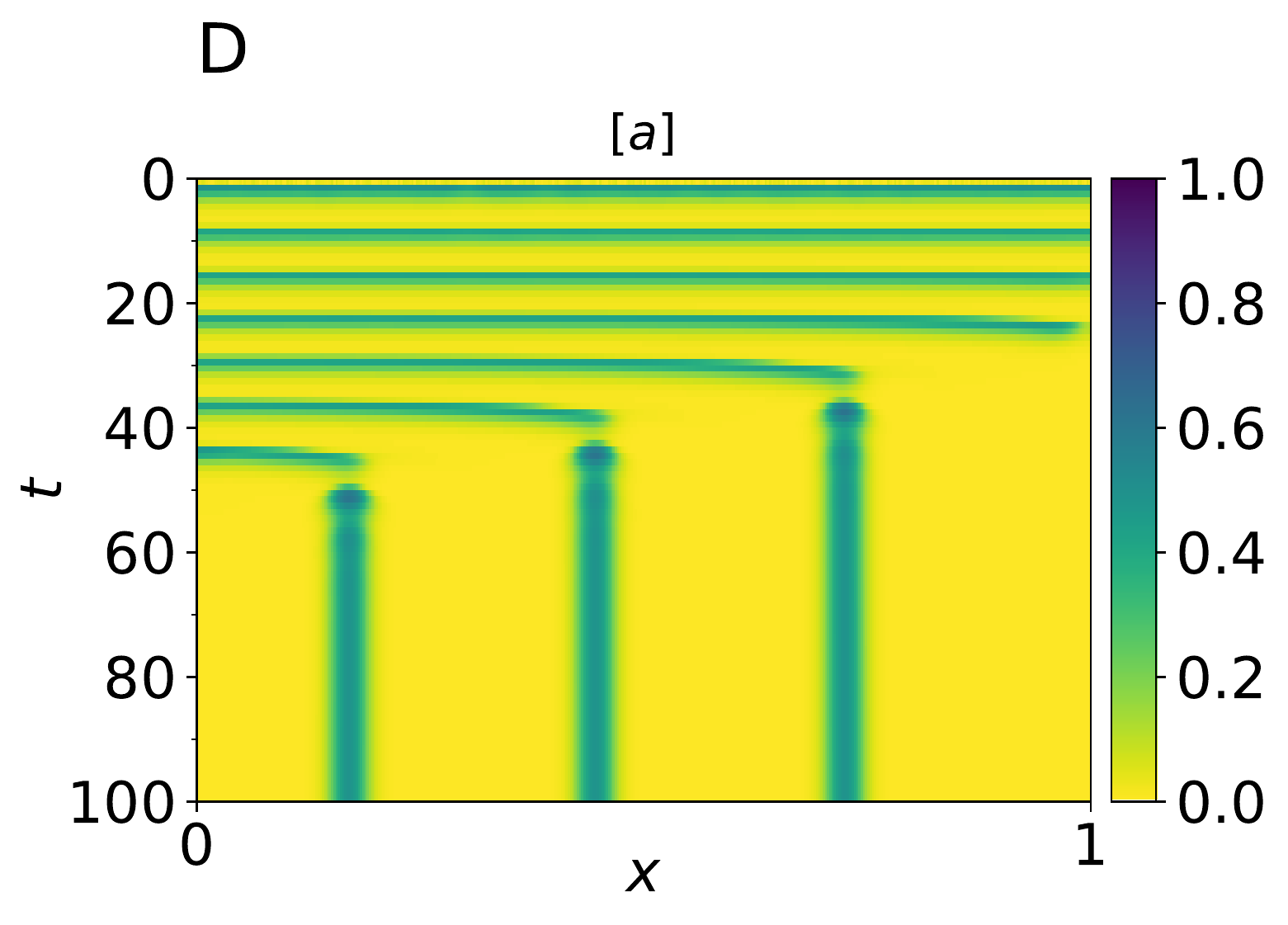}
		\caption{Spatio-temporal evolution of $a$ from system~\eqref{eq078910} for $\beta$-wave speed as in~\eqref{eqbeta}, with boundary condition as in~\eqref{eqbn}. (A) $\beta$-wave evolution and (B) somite-pattern formation for speed $v=0.02$; (C)~$\beta$-wave evolution and (D) somite-pattern formation for speed $v=0.04$. Initial conditions for~$a$ consists of random uniform distributions in the interval $[0, 0.1]$, whereas $r(x, 0) = 0$ for all $x\in[0,1]$. Parameter values were set as follows: $k_1=0.05$, $k_2=0.01$, $k_3=2$, $n=3$, $d_a = 5\times10^{-5}$ and~$d_r=2.5\times10^{-3}$. Animations with the same  data sets used to plot the heat-maps in this figure can be found in the reservoir \url{https://github.com/JesusPantoja/Reaction-Diffusion_Movies/}.}
		\label{Fig04}
	\end{figure}
	
To further disclose the above discussion, we analyzed the spatial stability of the PDE system when depending on parameters $\beta$ and $d_a$ in Appendix~\ref{app:bif}. We found that, when $d_a = 0$ and the system is equivalent to the Meinhardt-PORD model, the limit cycle does not turn spatially stable, even for quite large values of $\beta$. This makes it impossible to couple the gene network with a receding morphogen gradient and get a clock-and-wavefron behavior, as we have conceived it. Nonetheless, it is known that reaction-diffusion systems can stabilize because of diffusion. Thus, we wondered whether expanding the model by also accounting for diffusion of the activator would do the trick. As can be seen in Fig.~\ref{FigB01}, it is possible to stabilize the limit cycle by increasing the value of $\beta$, for $d_a$ values larger than about $d_a=2.5\times10^{-5}$; in other words, region III in Fig.~\ref{FigB01},  corresponding to sustained spatially stable limit cycles, is finite and subsists for $d_a>0$. To illustrate these findings, we present in Fig.~\ref{Fig03} the results of two simulations: one in which the limit cycle is stable and another in which it is unstable. 

\begin{figure}
	\centering
	\includegraphics[width=3in]{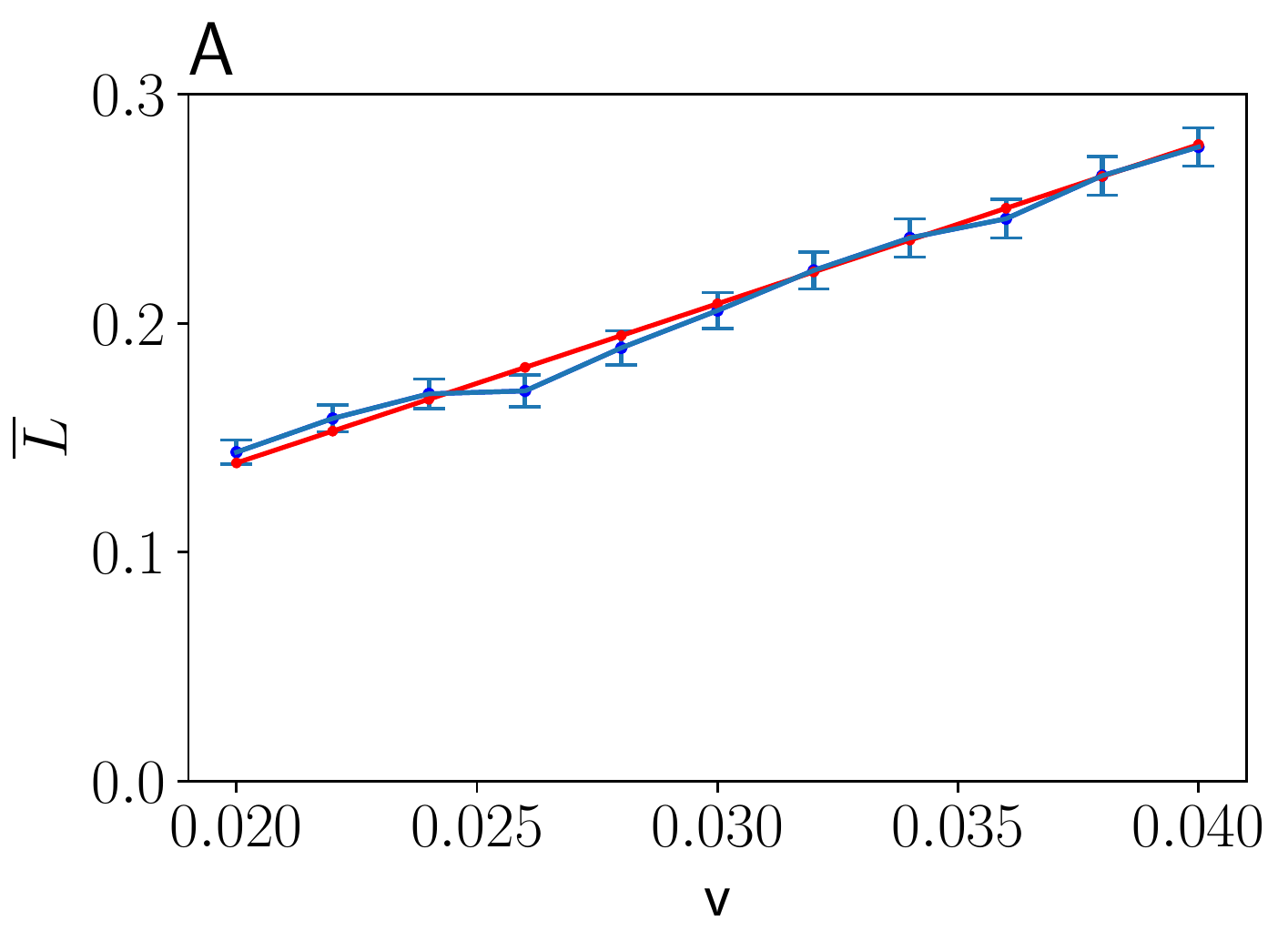} 
	\includegraphics[width=3in]{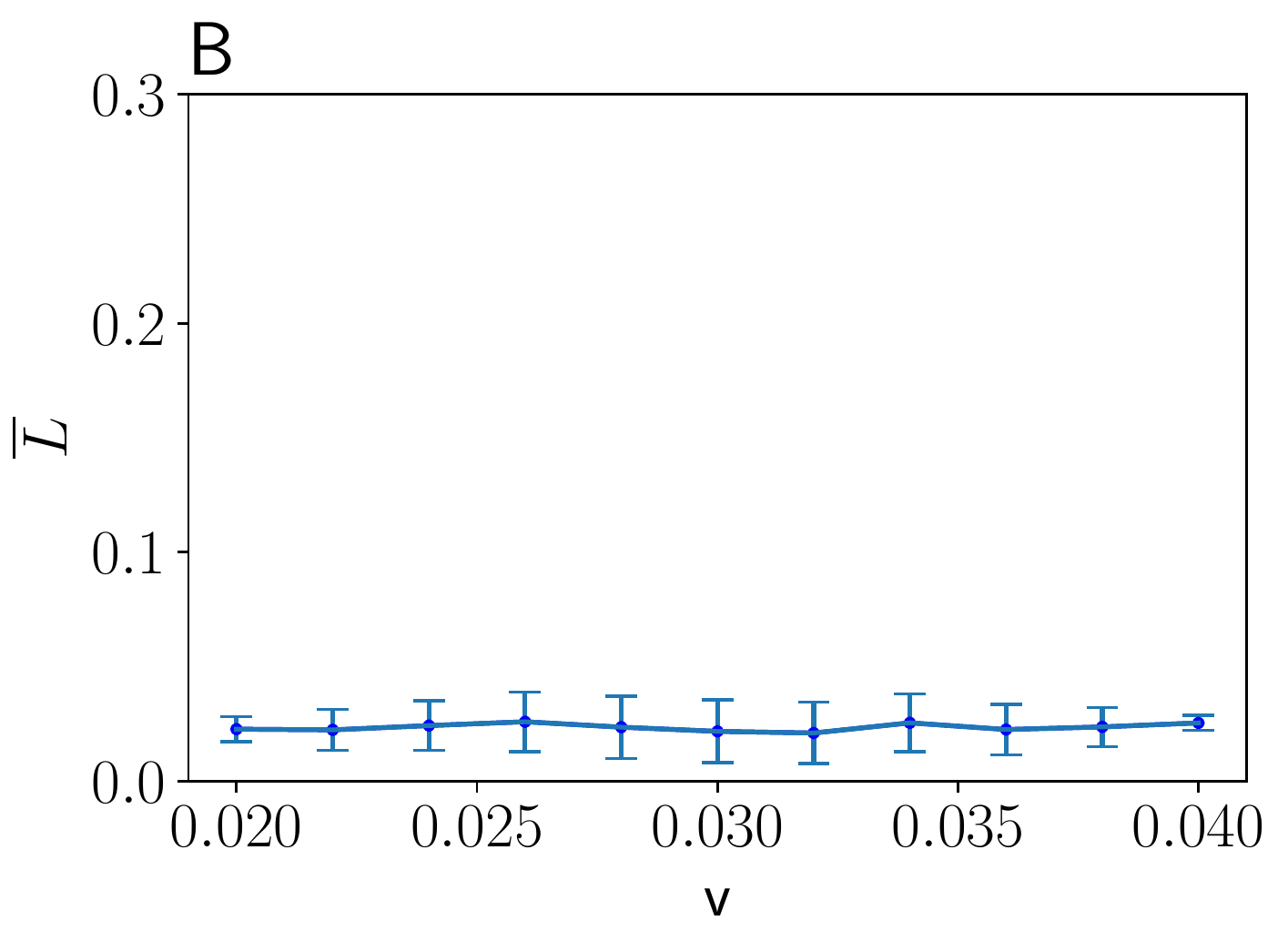}
	\caption{Plots of average somite size ($\overline{L}$) vs. velocity of the receding morphogen gradient ($v$) for two different cases: (A)~considering activator diffusion, $d_a = 5\times10^{-5}$ (in this case we employed random initial conditions), the red line corresponds to $\overline{L} = v T$, with $T$ the period of oscillations; and~(B) without activator diffusion, $d_a = 0$ (in this case we employed homogeneous initial conditions, except for a small perturbation at $x=1$). In all cases, the average somite size was computed from 10 independent simulations, and error bars denote standard deviation.}
	\label{Fig04_1}
\end{figure}

To test whether the bifurcations described in the previous paragraphs (when $d_a > 0$) are enough to couple the oscillatory gene network and a receding morphogen gradient, and yield a clock and wavefront behavior, we performed further simulations in which, instead of considering a constant value of~$\beta$, we assume that it is given by
	\begin{gather}\label{eqbeta}
	\beta(x, t) = \beta_0 \frac{K_{1/2}^m}{K_{1/2}^m + (x - v t)^m}\,.
	\end{gather}
Notice that this expression corresponds to a external regulation of gene A, which is a temporarily and spatially dependent profile that decays in a sigmoidal fashion in the posterior to anterior direction, and travels with speed $v>0$ in the opposite direction. That is, it mimics the behavior of the morphogen profile. The results of two such runs are shown in Fig.~\ref{Fig04}. In those simulations, we set $K_{1/2} =1.6 $, $m = 4 $ (these values were also employed for the simulation in Fig.~\ref{Fig05}), and considered two different speed values $v = 0.02 $, and $v = 0.04$. Observe that a periodic pattern arises in both simulations, with the same period as that of the oscillatory gene circuit. However, the regions corresponding somites are larger for the faster receding morphogen gradient. These results are consistent with a clock and wavefront mechanism, and agree with the experimental observation that somites are larger when the velocity of morphogen profile is increased \cite{Sawada2001}. In fact, we were able to corroborate that somite size is proportional to the wavefront velocity (see Fig. \ref{Fig04_1}A), an intrinsic feature of all systems in which somitogenesis is driven by an external wavefront interacting with a clock \cite{Morelli2009}. We can also appreciate  in Fig. \ref{Fig04_1}B that somite size is not regulated by velocity of the receding morphogen gradient when $d_a=0$, confirming that activator diffusion is necessary for an efficient coupling between the gene network and the receding morphogen gradient. Interestingly, when $d_a > 0$, we did not have to  assume a specific initial condition for the correct segmentation pattern to arise. The simulations in Figs.~\ref{Fig04} and \ref{Fig04_1}A were carried out considering random initial conditions, but the same results are obtained starting from uniform initial conditions. Conversely, we had to consider homogeneous initial distributions, except for a small perturbation at $x=1$, for the simulations in \ref{Fig04_1}B, which correspond to $d_a=0$.

The behavior depicted in Figs. \ref{Fig04} and \ref{Fig04_1}A is similar to several other models for somitogenesis, in which the external wavefront causes a transition from oscillation to bistability \cite{Francois2007}. What makes the present model different is the dynamic mechanism driving segmentation. In here, when parameter $\beta$ (which accounts for the interaction of the gene network with the external wavefront) locally decreases below a given threshold, the system dynamics turns spatially unstable. Hence, any local inhomogeneity is magnified and causes a biphasic pattern of gene expression. This mechanism has a couple of characteristics that make it appealing: a) since the system behavior changes when the value of $\beta$ is modified, a $\beta$ gradient gives rise to local inhomogeneities, and so no special initial condition is needed to seed patterning; and b) the fact that the system is spatially stable for large $\beta$ values, means that $\beta$ essentially introduces an external control of the instability, which in turn modulates the size of the emmerging pattern and attenuates naturally-occurring initial inhomogeneities.
	
We verified by performing several simulations (results not shown) that, in our model, the frequency of oscillations decreases together with the value of parameter $\beta$. This implies that, in simulations  with a receding morphogen gradient, anterior regions of the PSM oscillate at a lower frequency than posterior regions, in agreement with experimental observations \cite{Goldbeter2008, Hester2011}. This behavior can be appreciated in the simulations associated to Fig.~\ref{Fig04}, which can be found in the reservoir \url{https://github.com/JesusPantoja/Reaction-Diffusion_Movies/}. To the best of our knowledge, this an interesting result because we were able to get a frequency profile via a very simple model. As far as we know, most of the times it either has to be assumed a priori \cite{Morelli2009} or much more detailed models were necessary to reproduce it \cite{Hester2011}.
	
\begin{figure}
	\centering
	\includegraphics[width=3in]{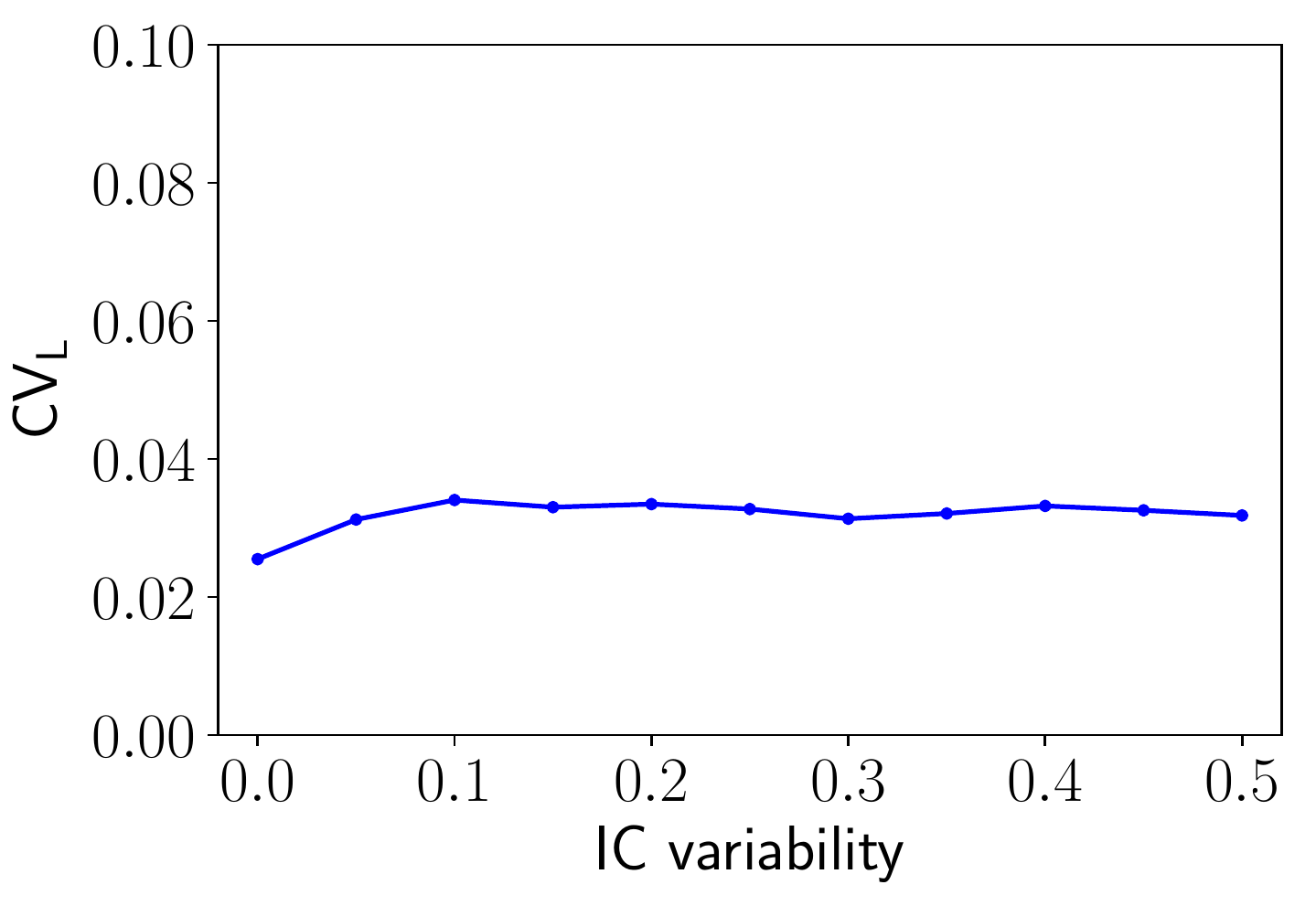}
	\caption{Plots of the coefficient of variation of somite size ($\text{CV}_L$) vs. the initial condition (IC) variability (measured as the size of the interval from which initial condition values were randomly selected).}
	\label{Fig06_1}
\end{figure}

\begin{figure}[t!]
		\centering
		\includegraphics[width=2in]{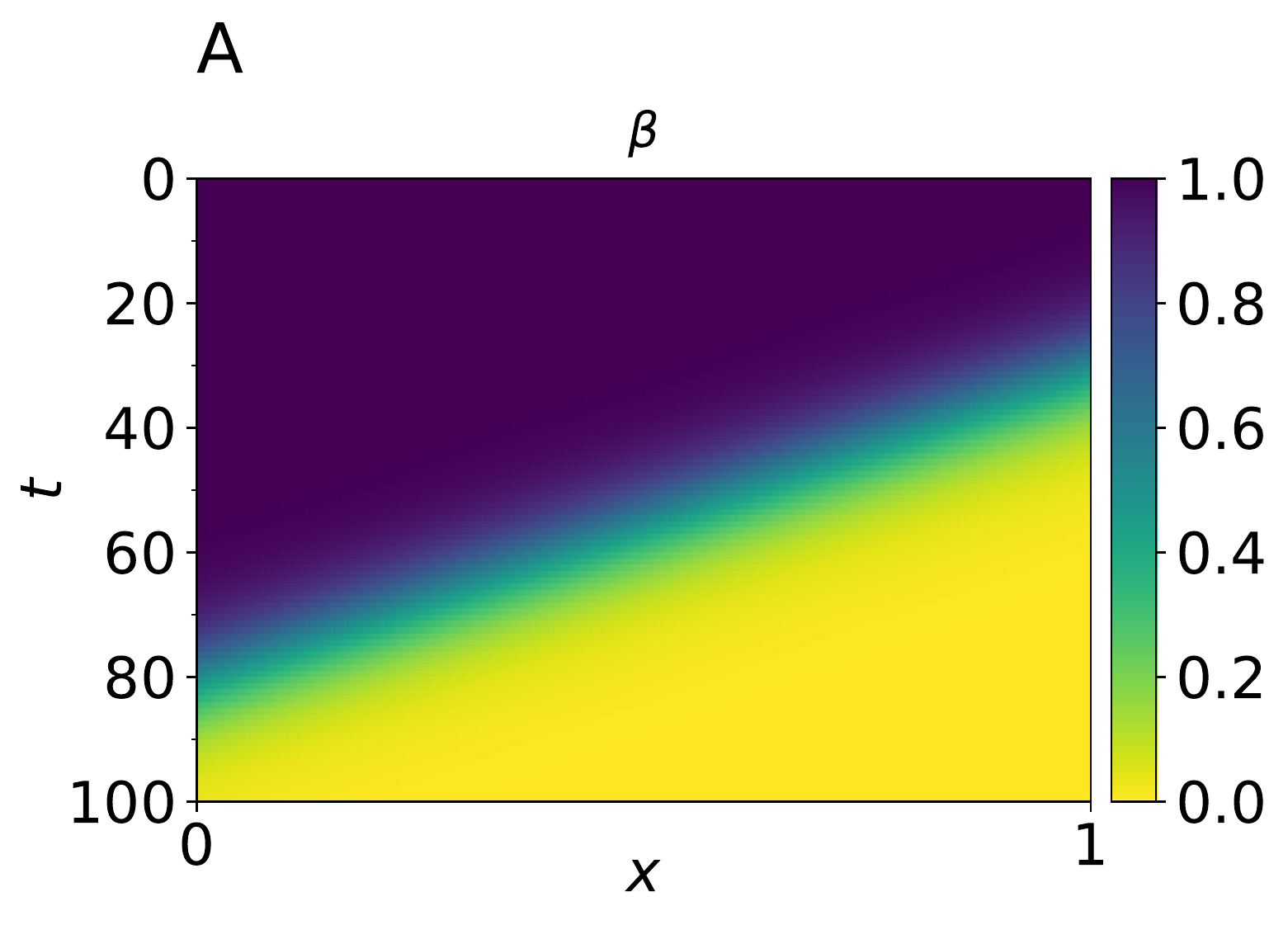}
		\includegraphics[width=2in]{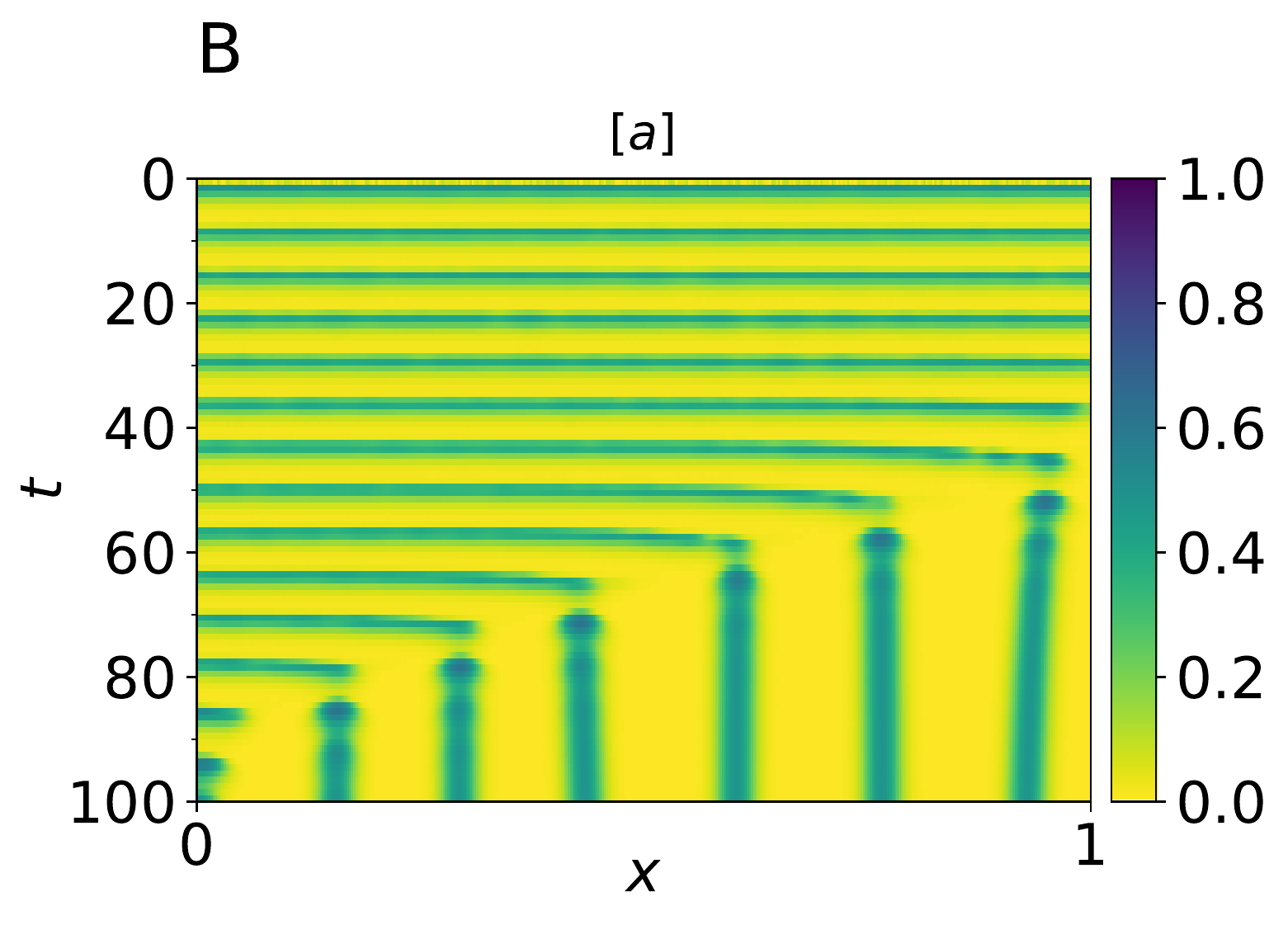} 
		\includegraphics[width=2in]{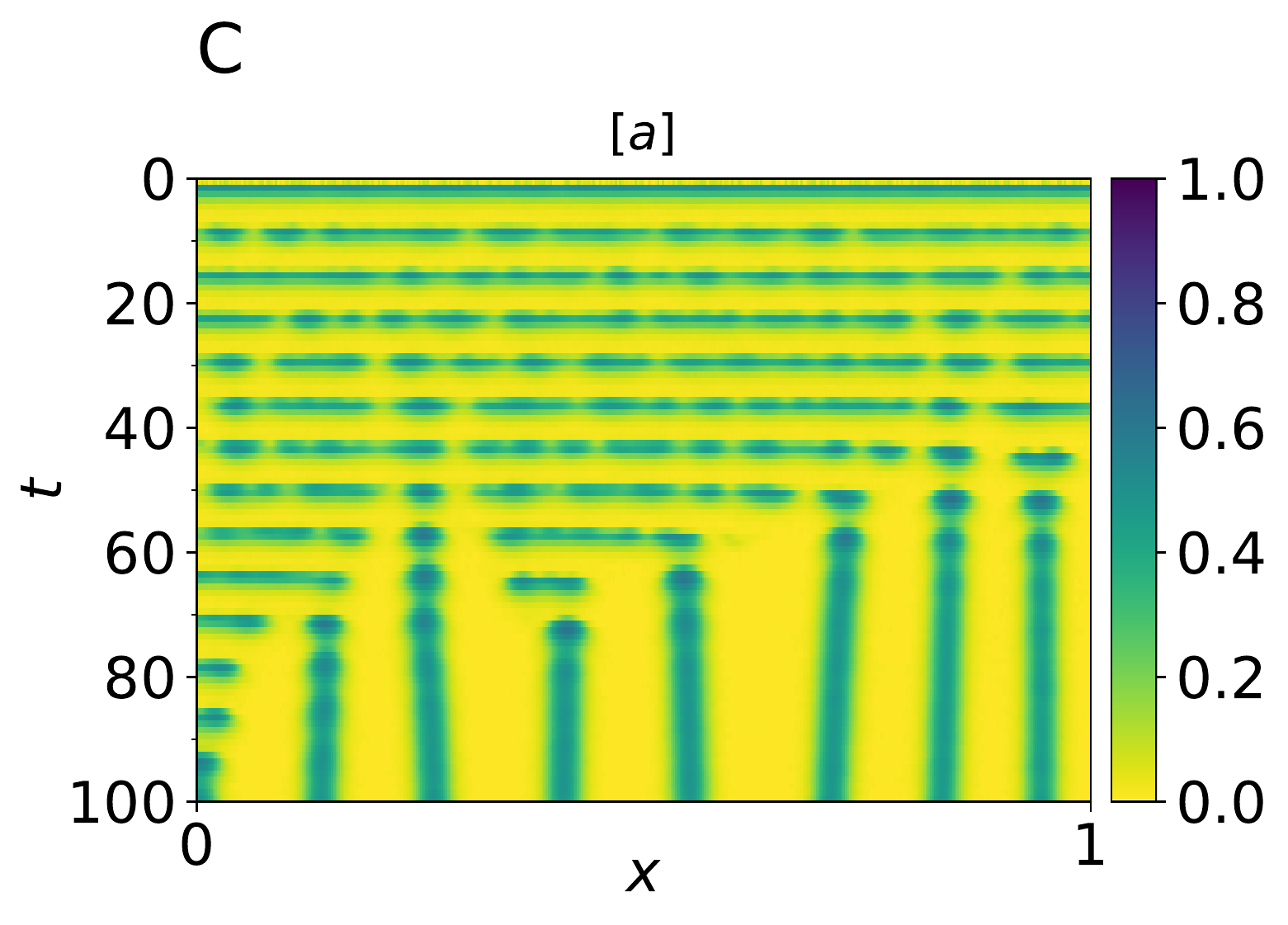} 
		\caption{Spatio-temporal evolution of $a$ from system~\eqref{eq1112} for $\beta$-wave speed as in~\eqref{eqbeta}, with boundary condition as in~\eqref{eqbn}: (A)~$\beta$-wave evolution for speed $v=8$ for somite-pattern formation with noise intensity (B)~$c_v=0.05$ and (C)~$c_v=0.1$. Initial conditions for $A$ consists of random uniform distributions in the interval $[0,0.1]$, whereas $R(x, 0) = 0$ for all $x\in[0,1]$. Parameter values were set as follows: $k_1=0.05$, $k_2=0.01$, $k_3=2$, $n=3$, $d_a = 5\times10^{-5}$ and~$d_r=2.5\times10^{-3}$. Animations with the same data sets used to plot the heat-maps in this figure can be found in the reservoir \url{https://github.com/JesusPantoja/Reaction-Diffusion_Movies/}.}
		\label{Fig05}
	\end{figure}

As earlier discussed, we expect that the present model behavior is robust to  of initial-condition variability, given that the system  oscillatory behavior is spatially stable at large $\beta$ values. To investigate this, we carried out several simulations in which the initial conditions where randomly selected from a uniform distribution in the interval $[0, m]$, with $m$ a parameter whose value we modified to change the level of initial condition (IC)  variability. The obtained results (not shown) confirmed our expectation at first sight. To quantify our observations, we carried out 10 independent simulations for each level of IC variability, and then computed the coefficient of variation of somite length $\text{CV}_L$, after pulling together the numerical data from the 10 simulations . The results are summarized in Fig. \ref{Fig06_1}. Observe that the $\text{CV}_L$ values are quite small in all cases, about 3\%, and do not depend on the IC variability level, thus confirming that the model dynamic behavior is extremely robust to variability of initial conditions. This issue is also discussed in a recent paper \cite{JutrasDub2020}, where a rather significant period gradient naturally emerges from a simple interplay between enhancers.

\begin{figure}
	\centering
	\includegraphics[width=3in]{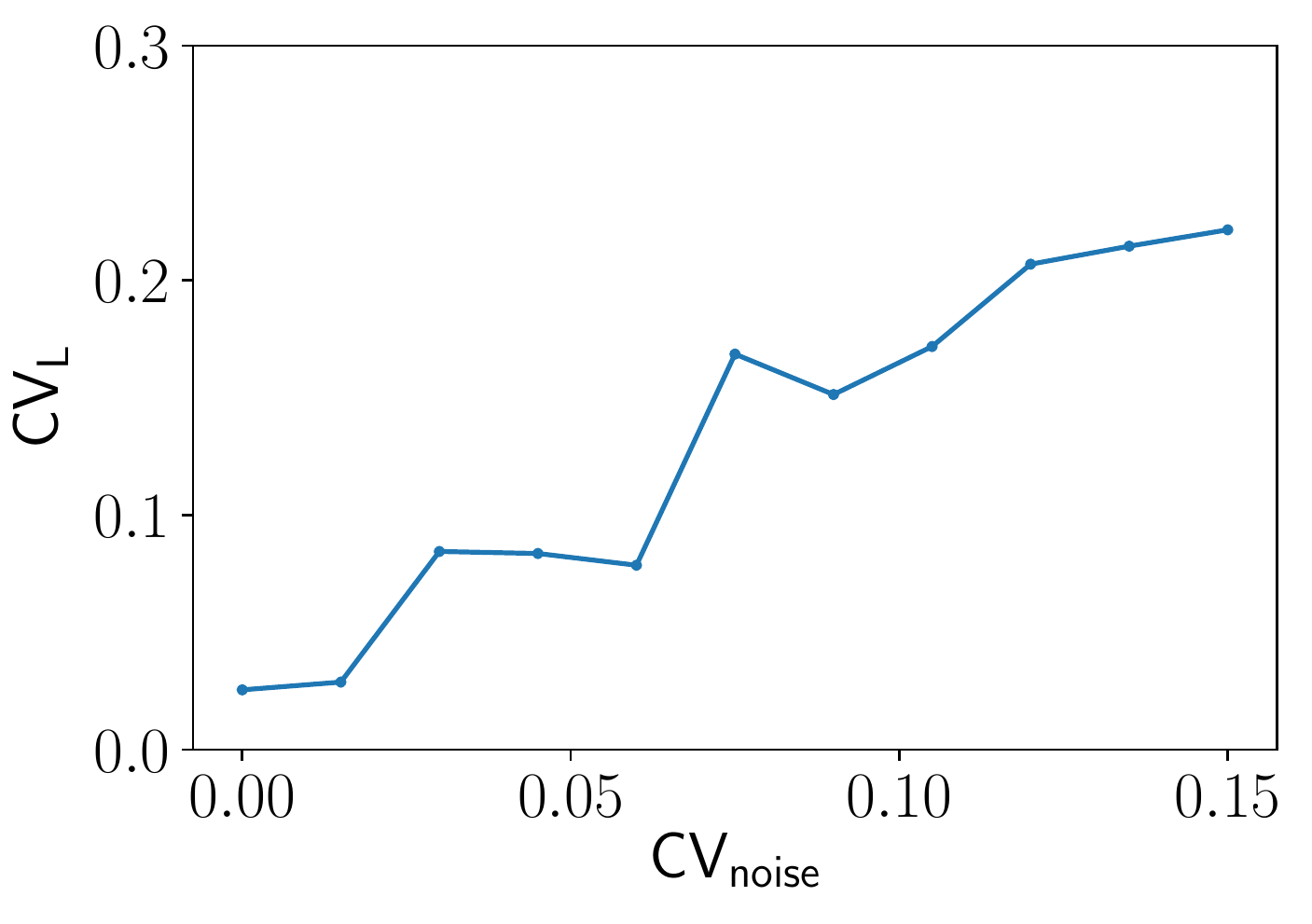}
	\caption{Plots of the coefficient of variation of somite size ($\text{CV}_L$) vs. the coefficient of variation of the added white Gaussian noise $\text{CV}_{\text{noise}}$.}
	\label{Fig06_2}
\end{figure}

We further tested the system robustness to added noise. To do so, we performed simulations in which white Gaussian noise was added to both variables ($a$ and $r$). In Fig.~\ref{Fig05}A, we show the result of a typical simulation in which the coefficient of variation of the added noise is $\text{CV}_{\text{noise}}= 0.05$ (relative to the corresponding variable steady-state value). Observe that somite formation proceeds in a precise way in this case. On the contrary, when $\text{CV}_{\text{noise}} = 0.1$ (Fig.~\ref{Fig05}B), although somites continue emerging, they do it in a disordered and non-periodical fashion. This happens because one of the effects of adding noise is altering the phase of the oscillating genes, and this modifies the timing of their interaction with the receding morphogen gradient. To quantify robustness to added noise, we performed numerous simulations, with different noise intensities (10 independent simulations for each noise intensity level), and measured the coefficient of variation of somite size. The results are shown in Fig. \ref{Fig06_2}. Notice that the variability of somite sizes is an increasing function the the intensity of added noise. This indicated that the system is not as robust to added noise as it is to variability of initial conditions, and can only proceed precisely with a moderate level of intrinsic and extrinsic noise. We speculate from this result that somitogenesis should have a mechanism to cope with noise in real life, and that this mechanism could be the the direct interaction and local synchronization of adjacent PSM cells \cite{PantojaHernndez2021}.

To summarize, the results presented in the above paragraphs confirm the hypothesis that, by properly modifying and coupling a Meinhardt-PORD model with an external wavefront (in the form of a receding morphogen gradient), it undergoes a bifurcation from a stable to an unstable limit cycle, as a result of its interaction with the external wavefront, and that this bifurcation is enough to explain clock and wavefront dynamics. To the best of our understanding, this hybrid model has a couple of quintessential features: (i)~it explains why somitogenesis may occur in the absence of an external wavefront (albeit in an irregular fashion), and (ii)~recovers all of the features of clock and wavefront models. In particular, it accounts for the robustness of somitogenesis to random variations in the initial conditions.

	\section{Concluding Remarks}
	\label{conclu}
	
We have expanded an equivalent Meinhardt-PORD model (by assuming that not only the repressor but also the activator diffuse) and coupled it with a receding morphogen gradient (such as FGF8 in chicken and mice), to have a hybrid model for somitogenesis, that incorporates characteristics of the Meinhardt-PORD and clock-and-wavefront models.  This hybrid model undergoes a bifurcation, from a stable to an unstable limit cycle, as the value of the parameter accounting for the background regulatory input of the activator decreases. From a biological perspective, the bifurcation just described allows the model to explain why somites can form in the absence of an external wavefront (which traditional front and wavefront models failed to explain), reassesses the role of the receding morphogen gradient as a conductor for somitogenesis, and makes the model behavior robust to initial condition variability, as well as independent from specific initial conditions; notice that the latter are two of the weak points of the original PORD model. 
	
	In the clock and wavefront models, there is a consensus that the oscillatory behavior is originated by a gene network with time-delayed negative feedback regulation \cite{Monk2003, Lewis2003}. This claim is supported by multiple experimental reports, which have elucidated some of the underlying regulatory mechanisms \cite{Schroter2012}. In contrast, the present model (together with Meinhardt-PORD models) not only relies on a different gene network architecture to generate oscillations, but the genes in the network are hypothetical. From this perspective, the weight of experimental evidence seems to favor clock and wavefront models. Nonetheless, upon taking this into consideration, we believe that we have provided convincing evidence that reaction-diffusion and positional information (receding morphogen gradient) mechanisms could work together in somitogenesis. Further investigating this possibility may allow a better understanding of such a fascinating phenomenon.

\section*{Acknowledgements}

The authors are grateful to the anonymous reviewers, whose comments and criticisms greatly helped us to improve our manuscript. JP-H thanks CONACYT-México for granting him a doctoral scholarship. VFBM thanks the financial support by Asociación Mexicana de Cultura AC. MS acknowledges the financial support of CONACYT-México, grants INFRA-19-302610 and CDF-19-568462. 

\section*{Data availability statement}

The data that support the findings of this study are available from the corresponding author upon reasonable request.
	
	
	\appendix
	
	\section*{Appendix}
	\section{Somite pattern-formation dynamical features}
	\label{app:bif}

	\begin{figure} 
		\centering
		\includegraphics[width=3in]{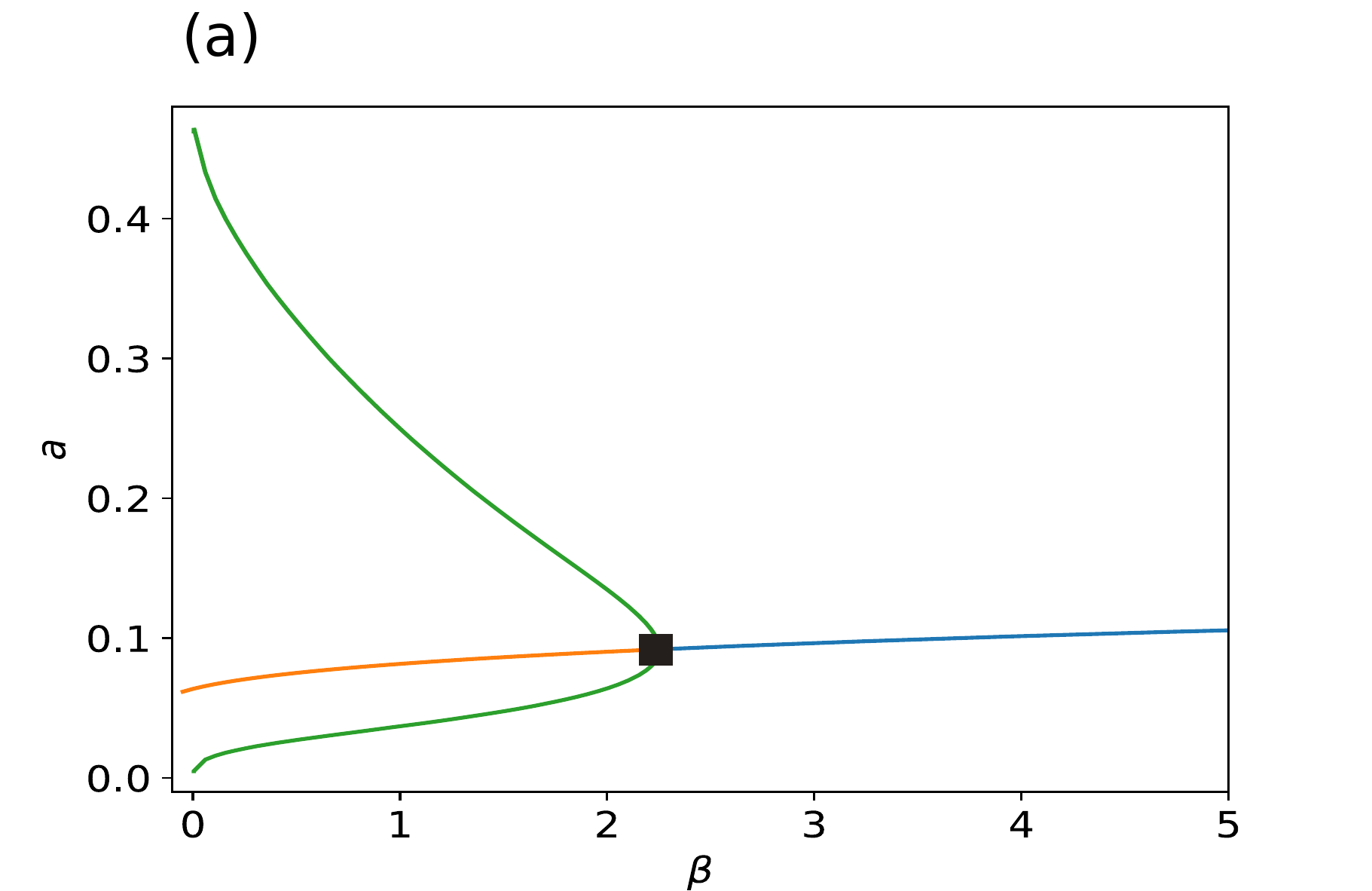} \\
		\includegraphics[width=2in]{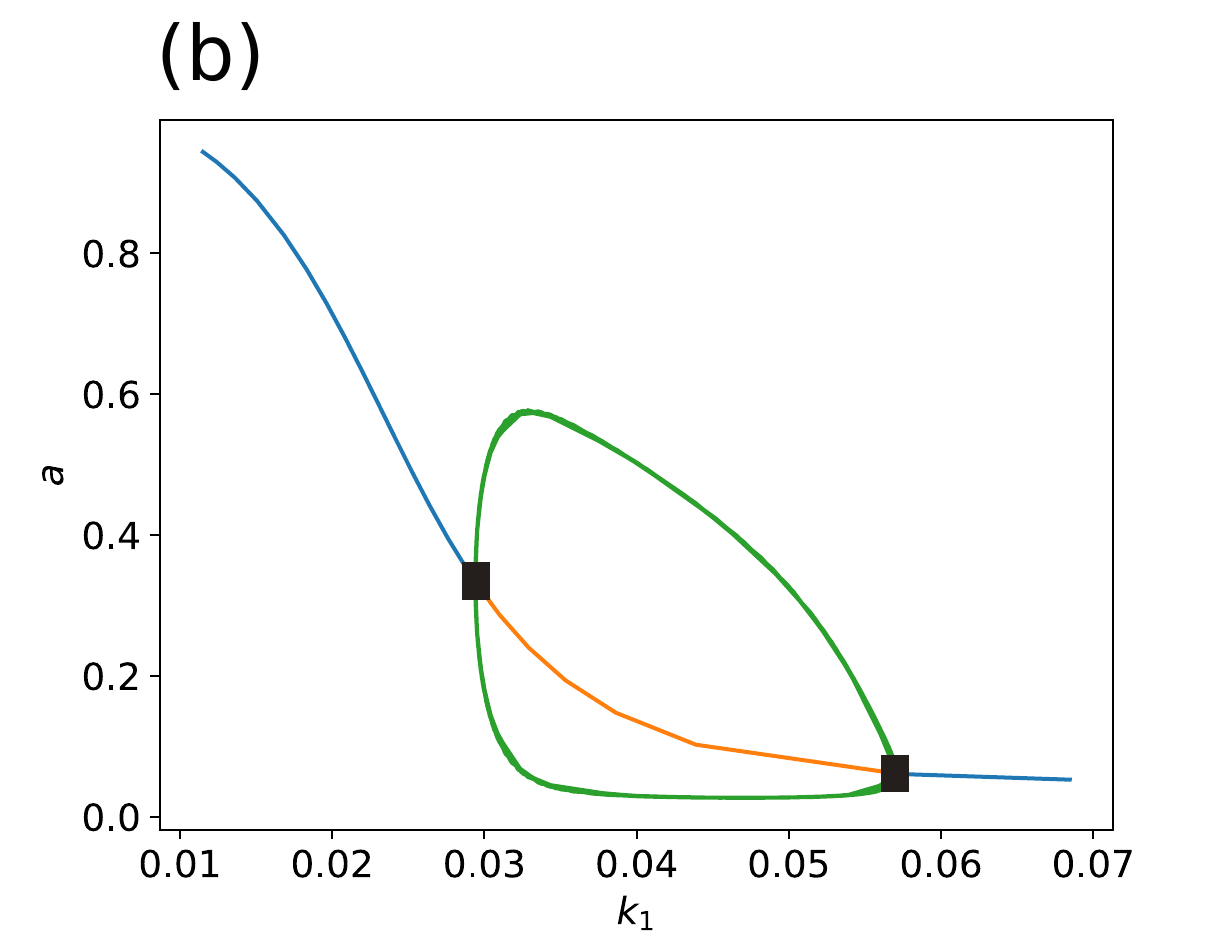}
		\includegraphics[width=2in]{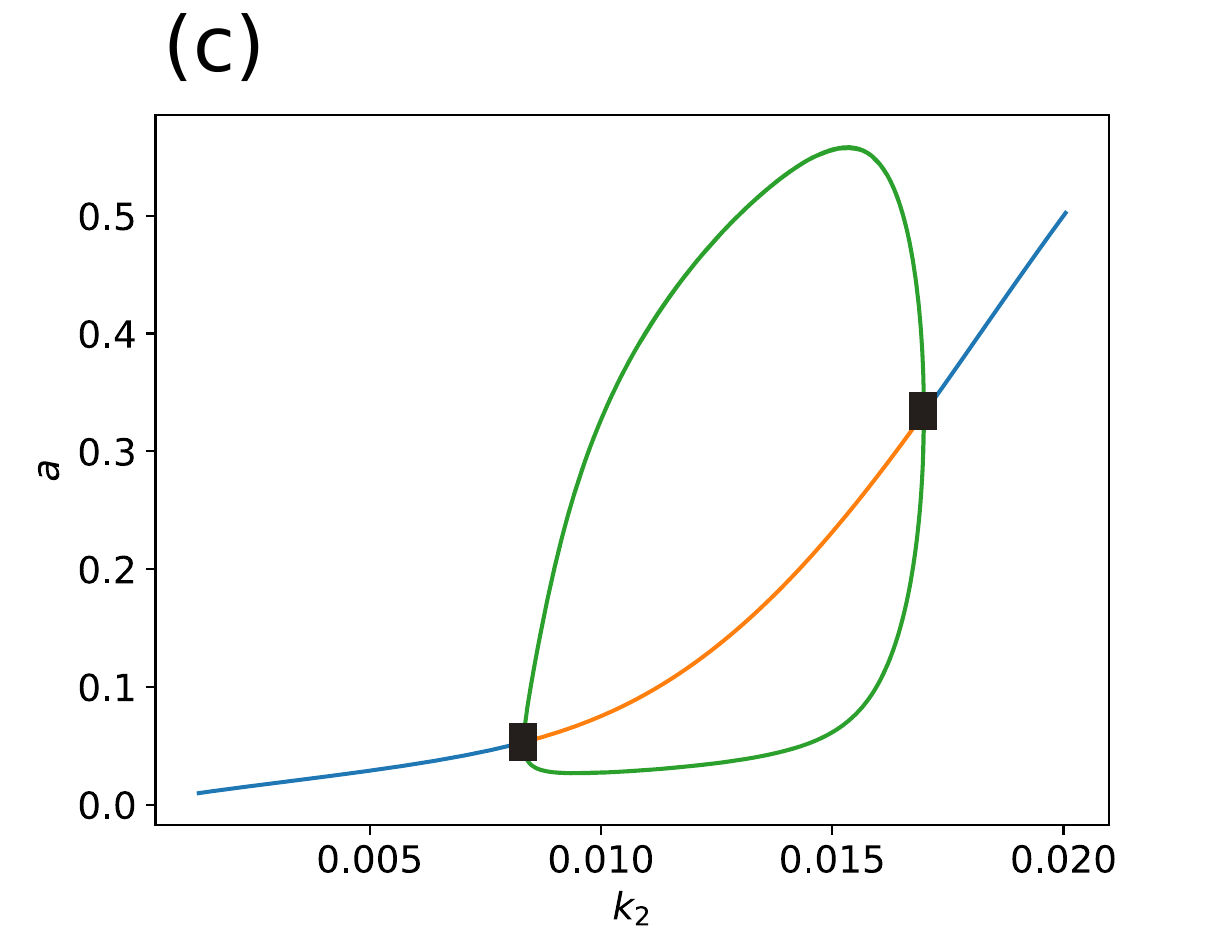} \\
		\includegraphics[width=2in]{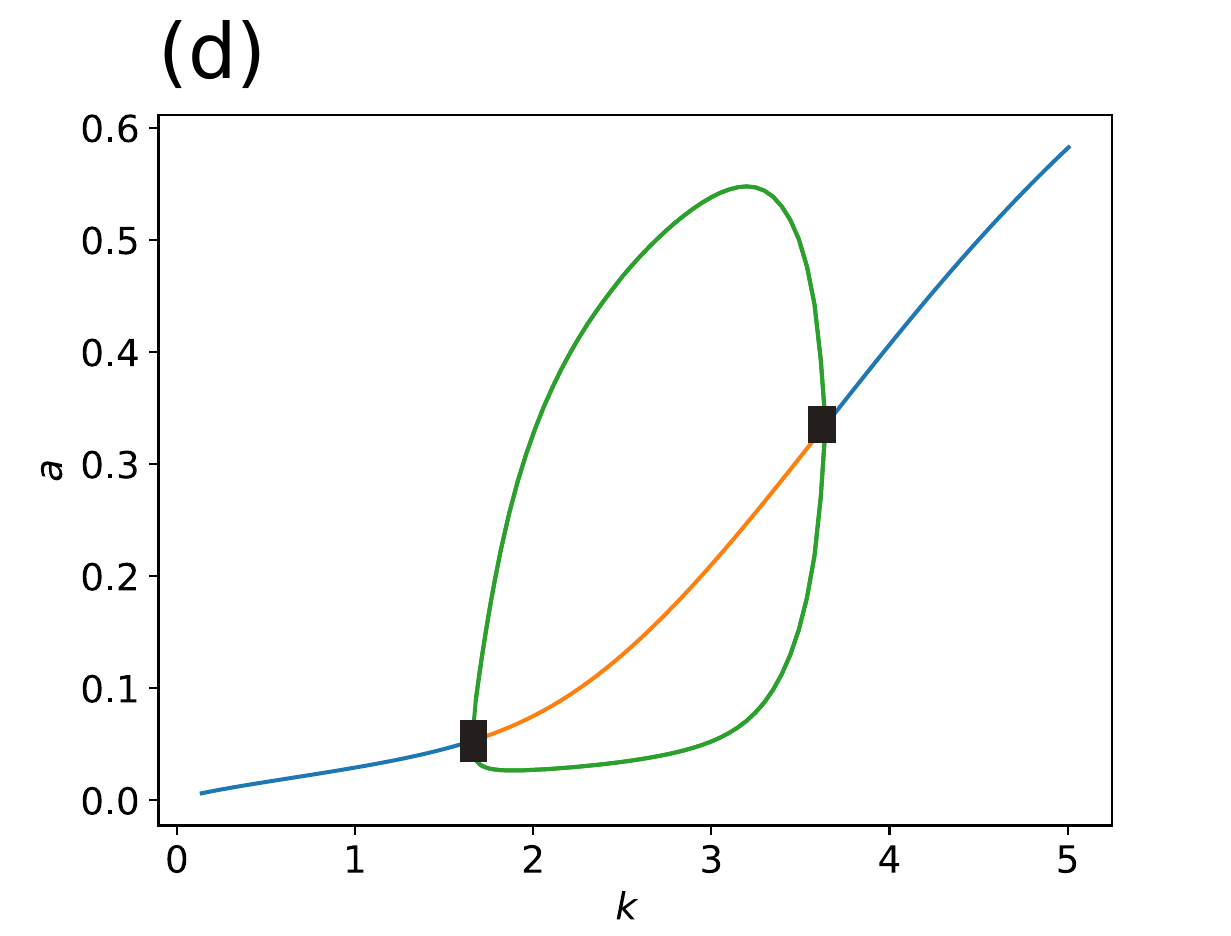}
		\includegraphics[width=2in]{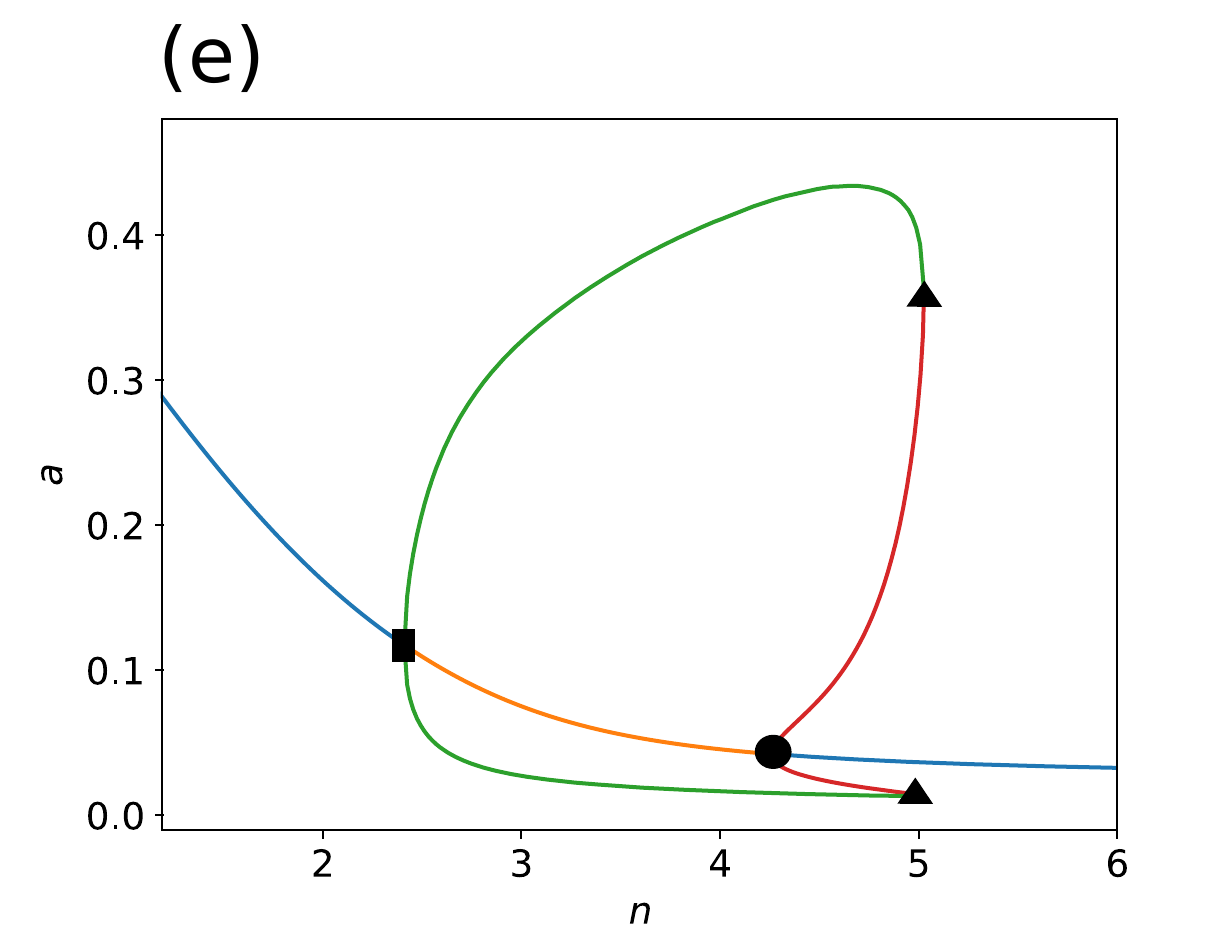}
		\caption{Bifurcation diagrams of system~\eqref{eqA12} for slowly varying: (a)~background regulatory input of activator $\beta$, effective half-saturation parameters (b)~$k_1$, (c)~$k_2$ and (d)~$k_3$, and (e)~Hill coefficient~$n$. Blue (orange) solid lines correspond to stable (unstable) branches of steady states, and green (red) lines are stable (unstable) limit-cycle branches. Squares (filled circle) indicate supercritical (subcritical) Hopf bifurcations, and triangle is for indicating fold bifurcations. Other parameter set values are $k_1 = 0.05$, $k_2 = 0.01$, $k_3 = 2.0$, $n=3.0$, and $\beta=0.5$, respectively.}
		\label{FigA01}
	\end{figure}

	System~\eqref{eq078910}, along with homogenous Neumann boundary conditions, gathers the essential ingredients of somite pattern-formation dynamics. Particularly, the kinetic terms consist of Hill functions, whose coefficients corresponding to a non-negative cooperative interaction. Thus, we assume that $n_3 = 1$ and $n_1 = n_2 = n\geq1$ as well as $\mu=1$. We start by analyzing the temporal behavior of the homogeneous system. To this end, we set $d_a=d_r=0$. In so doing, we get the purely kinetic system:
	\begin{subequations}\label{eqA12}
		\begin{gather}\label{eqAR12}
		\frac{da}{dt} = f(a,r)\,, \quad 
		\dfrac{dr}{dt} = g(a,r)\,, 
		\end{gather}
		where the field components are given by
		\begin{gather}\label{eqnB034}
		f(a,r) = \dfrac{\beta + (a / k_1)^{n}}{1 + (a / k_1)^{n}
			+ (r / k_2)^{n}}-a\,, \quad g(a,r) = \dfrac{a}{k_3 + a}-r\,. 
		\end{gather}
	\end{subequations}
	This system steady states satisfy the relation $\mathcal{H}(a)=\beta$, where
	\begin{gather}\label{eq:ga}
		\mathcal{H}(a):=a-a^n/k_1^n+a^{n+1}/k_1^n +a\left(\frac{a/k_2}{k_3+a}\right)^n\,.
	\end{gather}
	From~\eqref{eq:ga}, notice that: (i)~$\mathcal{H}(0) = 0$, and (ii)~$\mathcal{H}(a)\rightarrow$ +$\infty$, when $a\rightarrow +\infty$. In consequence, there exists $a^*>0$ such that $\mathcal{H}(a^*) = \beta> 0 $, which further implies that $w^*=(a^*, r^*)$ is a steady-state of system~\eqref{eqA12}, with $a^* >0$ and $r^*>0$, since 
	\begin{gather*}
	r^* = \frac{a^*}{k_3+a^*}\,.
	\end{gather*}
	In other words, the existence of at least one steady-state in the first quadrant is guaranteed. We can straightforwardly prove, from the Poincare-Bendixon theorem, that a limit-cycle family emerges as a result of this steady-state undergoing a Hopf bifurcation (HB). In order to disclose this implication, we performed a numerical continuation by using the algorithms implemented in XPPAUT (Ermentrout, 1987). The resulting bifurcation diagrams obtained by slowly varying each parameter of system~\eqref{eqA12} are depicted in Fig.~\ref{FigA01}. As can be seen there, no co-existing steady-states were found, and the non-spatially extended system~\eqref{eqA12} undergoes HBs for the parameters values in Table~\ref{Tab01}. In Fig.~\ref{FigA01}(a), a single supercritical HB for parameter $\beta$ takes place, where periodic orbits vanish at the bifurcation point HB. This indicates that, whenever no diffusion transport  is present, no unstable orbits exists. Additionally, observe that the lower the $\beta$-input value, the larger the amplitude and the longer period of stable orbits are obtained; and that no periodic orbits occur for $\beta\gg1$, nonetheless. In contrast, as is shown in Figs.~\ref{FigA01}(b)-(d), two supercritical HB points occur for parameters $k_1$, $k_2$ and $k_3$, which determine a finite interval for each parameter wherein a family of periodic orbits exist. Interestingly, a bi-stability interval for parameter~$n$ is delimited by $n$-values where a subcritical HB and fold bifurcation (LP) points happen. That is, a branch consisting of unstable limit cycles emanates from the subcritical HB, which stabilizes at the LP points. Hence, stable steady-states coexist with stable limit-cycles for this interval, where the unstable periodic branch plays a critical role for initial states. This result, as is depicted in Fig.~\ref{FigA01}(e), indicates that an on-and-off gene switch is present, which provides hysteretical features triggered by key values of the characteristic Hill coefficient. We may argue from this that the cooperativity in the gene regulatory network favors system robustness to parameter variations.
	
%
	
We turn now our attention to the spatio-temporal system dynamics. As has been discussed above, the external regulation of gene $A$ allows us to couple the gene network with a receding morphogen gradient. Recall that this element is captured by $\beta$. In addition, we have revealed that diffusion of the activator, which is characterized by $d_a>0$, is crucial to get an efficient coupling in our formulation. Hence, we now focus our analysis to the understanding of the interplay between $\beta$ and $d_a$, which triggers the somite formation mechanism that we have proposed. To do so, we include diffusion terms in~\eqref{eqA12} to get the reaction-diffusion system
		\begin{gather}\label{eqnB01234}
		\frac{\partial{a}}{\partial{t}}= f(a,r)+d_a\nabla^2a\,, \quad \frac{\partial{r}}{\partial{t}} = g(a,r)+d_r\nabla^2r\,,
		\end{gather}
		where the kinetic terms are as in~\eqref{eqnB034}. 

First of all, biologically plausible steady-state solutions are provided by real non-negative roots of the kinetic field for system~\eqref{eqnB01234}. Upon defining $w=(a,r)^T$, we obtain that system~\eqref{eqnB01234} can be set up in vector notation as $w_t = F(w) + D \nabla^2 w$, where
	$F(w) = \left(f(a,r),g(a,r)\right)^T$ and $D=\textrm{diag}\left(d_a, d_r \right)$. In so doing, for an isolated root of $F(w)=0$ given by $w^*=(a^*,r^*)$, where $a^*,r^*>0$, system~\eqref{eqnB01234} has a local solution of the form 
	\begin{gather}\label{eqnB06}
	w(x,t)= \sum_{m=0}^{\infty}\gamma_me^{-\lambda\left(\kappa^2\right) t}w_m(x)\,,
	\end{gather}
	where $w_m(x)$ satisfies the Helmholtz equation $\nabla^2w_m+\kappa^2w_m=0$, in which the so-called wave mode is denoted by $\kappa$. The Fourier coefficients $\gamma_m$ are determined by the initial conditions, and $\lambda$ determines whether~\eqref{eqnB06} converges, and hence is bounded, as $t\to+\infty$. These three parameters not only shape solution~\eqref{eqnB06}, but also are intrinsic to the geometry and boundary conditions of the system into consideration. Moreover, they also depend on the wave number $m\in\mathbb{N}$; see~\cite{Murray1989} for further details. 

\begin{figure}[t!]
	\centering
	\includegraphics[width=5.0in]{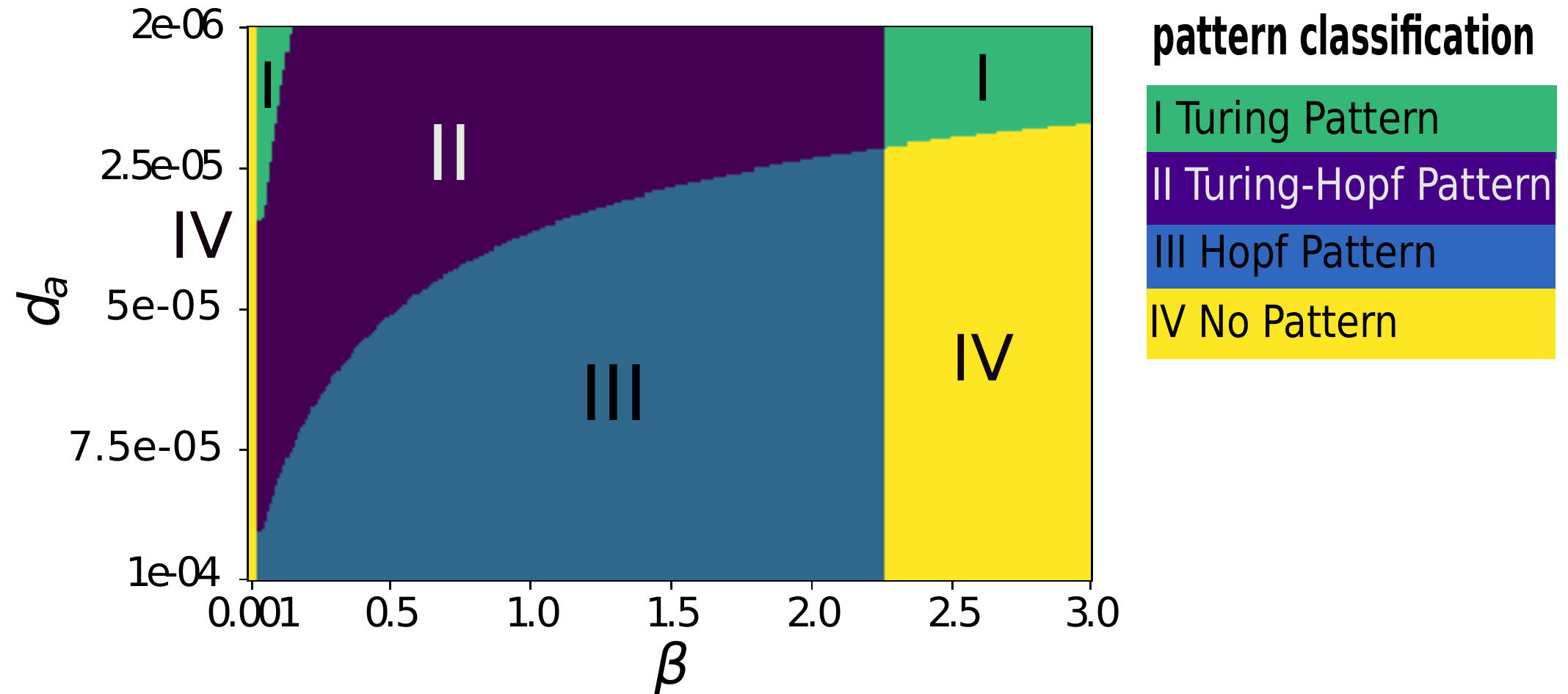}
	\caption{Two-parameter space for $d_a$ and $\beta$. Each pattern region is plotted in colors accordingly to the table in the right-hand side, and transition lines correspond to each region boundary. Other parameter set values are $d_r = 2.5\times10^{-3}, k_1 = 5\times10^{-2}, k_2=10^{-2}, k_3=2.0, n=3.0$.}
	\label{FigB01}
\end{figure}

We now derive the dispersion relation, which gives a linear insight of solution features depending on the parameters. To do so, we linearize system~\eqref{eqnB01234} at $w^*$ to get, in vectorial form,
	\begin{gather}\label{eqnB07}
	w_t = Jw + D \nabla^2 w	\,,
	\end{gather}
	where $J$ is the Jacobian matrix at $w^*$. Now, as our interest lies on the dynamics in one spatial dimension, we have that $w_m(x)=\cos(\kappa x)$, where $\kappa=m\pi/L$, satisfies the Helmholtz equation for homogenous Neumann boundary conditions as in~\eqref{eqbn}. Thus, \eqref{eqnB07} is satisfied by~\eqref{eqnB06}, when the dispersion relation $\lambda=\lambda(\kappa)$ is given by
	\begin{gather}\label{eqnB13}
	|J-D\kappa^2-\lambda I|=0\,, \quad I\in\mathbb{R}^{2\times2}\,,
	\end{gather}
	which can be seen by substituting~\eqref{eqnB06} into~\eqref{eqnB07}. As a result, it relates the temporal growth rate and the spatial wave mode $\kappa$, which parametrises the finite spatial domain. Note that~\eqref{eqnB13} leads to 
	\begin{subequations}\label{eq:lambda}
		\begin{gather}\label{eqnB14}
		\lambda^2 + b(\kappa^2)\lambda + c(\kappa^2) =0\,,
		\end{gather}
		where
		\begin{flalign}
		& b(\kappa^2)=(d_a+d_r)\kappa^2 -(f_a+g_r)\,, \label{eqnB15} \\
		& c(\kappa^2)=d_ad_r\kappa^4-(d_rf_a+d_ag_r)\kappa^2+ (f_ag_r-f_rg_a) \,.\label{eqnB16}
		\end{flalign}
	\end{subequations}
As we are interested in the linear stability of the steady-state $w^*$, characterized by parameters~$\beta$~and~$d_a$, we notice that the parameter space for spatial instability of Turing type is given by conditions: 
\begin{subequations} \label{eqnB17}
	\begin{gather}
	f_a+g_r<0\,, \quad f_ag_v-f_rg_a>0 \,, \label{eqnB18} \\
	d_rf_a + d_ag_r > 0 \,, \quad d_rf_a + d_ag_r > 2\sqrt{d_ad_r(f_ag_r-f_rg_a)} \,. \label{eqnB20}
	\end{gather}
\end{subequations} 
These conditions provide the ingredients that give place to non homogeneous spatially extended stationary states. We are however interested in a mechanism that triggers sustained oscillations of the gene network that give place to a stationary pattern in the long term. Such a process is dynamically provided by the Turing-Hopf bifurcation (THB); see, for instance,~\cite{Castillo2016, liu}. From~\eqref{eqnB14}, this bifurcation is prompted by obtaining purely imaginary eigenvalues by slowly varying $d_a$ or $\beta$. In so doing, we notice that two key conditions must be met: (i)~$f_a + g_r=0$ at the THB point, and (ii)~$d\lambda(\kappa^2;p^*)/dp\neq0$, also known as transversality condition, where $p^*=d_a^*$ or $p^*=\beta^*$ represents the THB parameter value. Thus, in order to obtain the parameter regions where a Turing bifurcation, HB and THB occur in system~\eqref{eqnB01234}, we solve~\eqref{eq:lambda} for $\lambda(\kappa^2)$. In Fig.~\ref{FigB01}, the parameter space on scope is portrayed, where four different stability features for the selected range values of parameters~$\beta$ and~$d_a$ are identified in four regions, labeled accordingly: Turing pattern, Turing-Hopf pattern, Hopf pattern, and no pattern. Notice that, at the transition curve between region II and III, parameters $\beta$ and $d_a$ follow an inverse relation; in other words, the larger parameter~$\beta$ is, the lower diffusivity~$d_a$ is needed for the bifurcation to take place. Nonetheless, for parameter values of $\beta$ large enough, this transition does not occur as no oscillating behaviors persist. In addition, notice that for a fixed value of $d_a=5\times10^{-5}$, slowly varying $\beta$ from 0 up to 2.25 drives the system through two bifurcations, which in turn originate three different mechanisms. That is, no pattern is formed for $0\leq\beta < 0.5\times10^{-2}$, which is followed by getting into the Turing-Hopf pattern region for $ 0.5\times10^{-2}<\beta<0.5$, to get in the Hopf pattern region, which is held by $ 0.5<\beta< 2.25$.

\begin{figure}[t!]
	\includegraphics[width=3.2in]{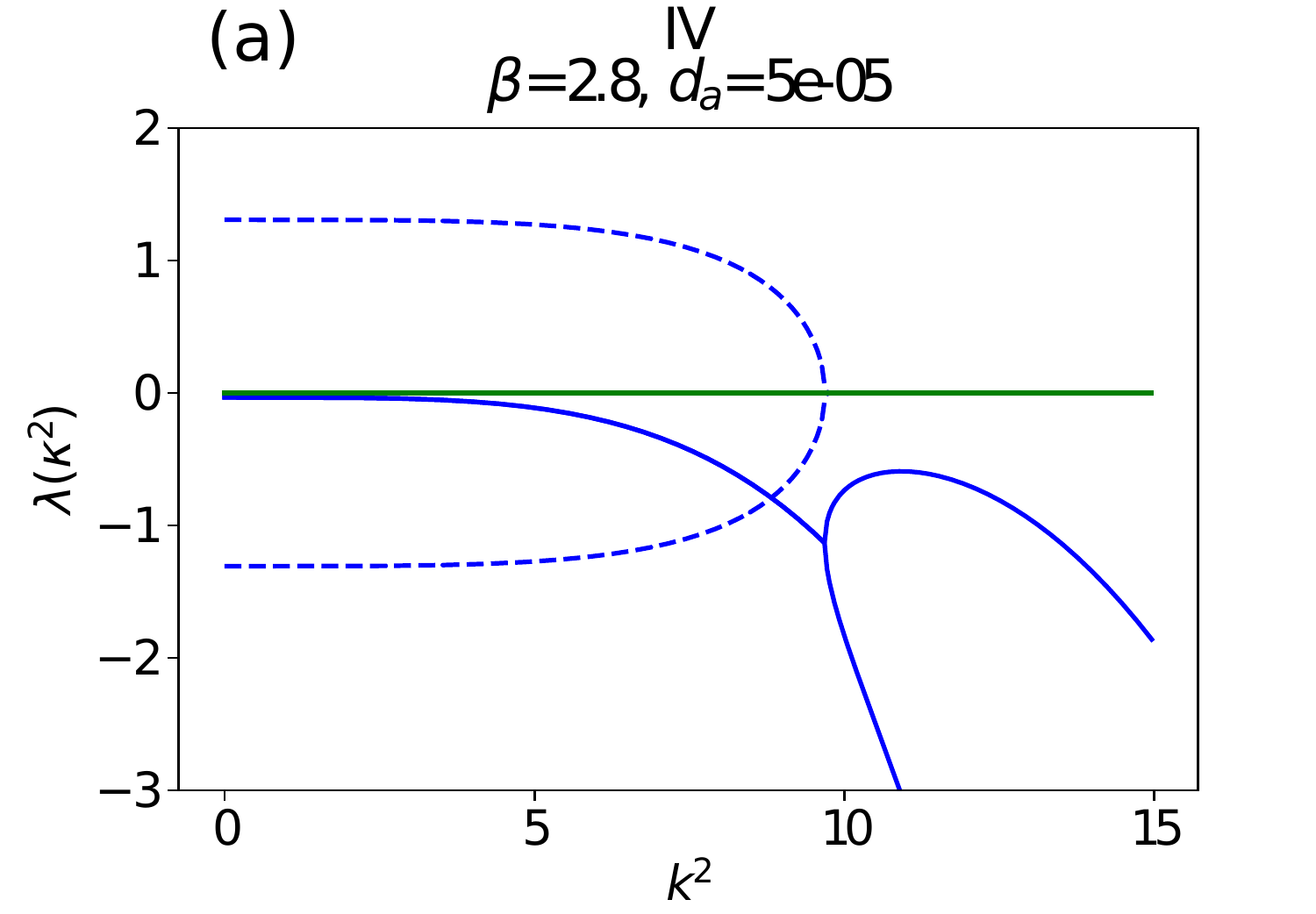} 
	\includegraphics[width=3.2in]{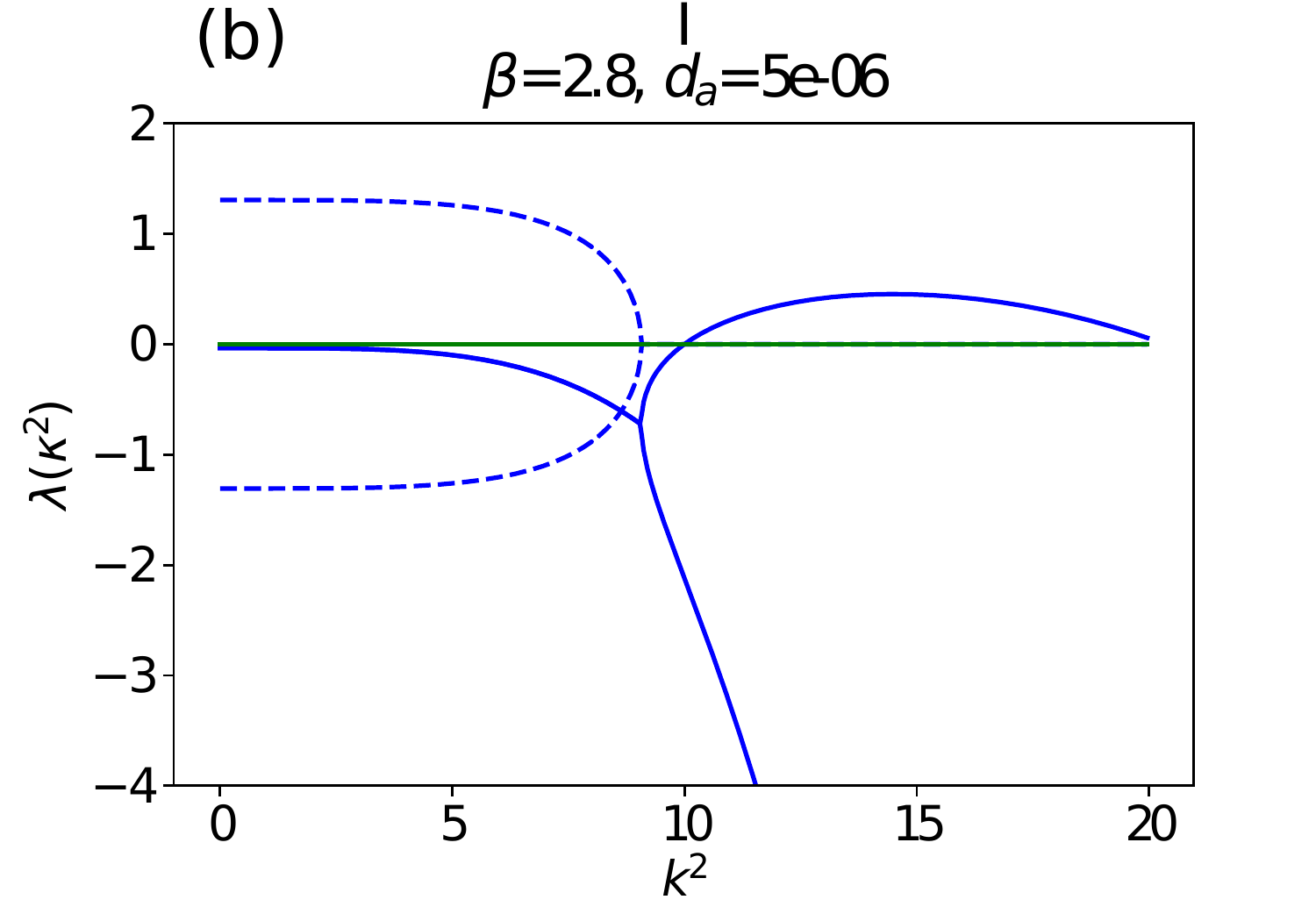}\\
	\includegraphics[width=3.2in]{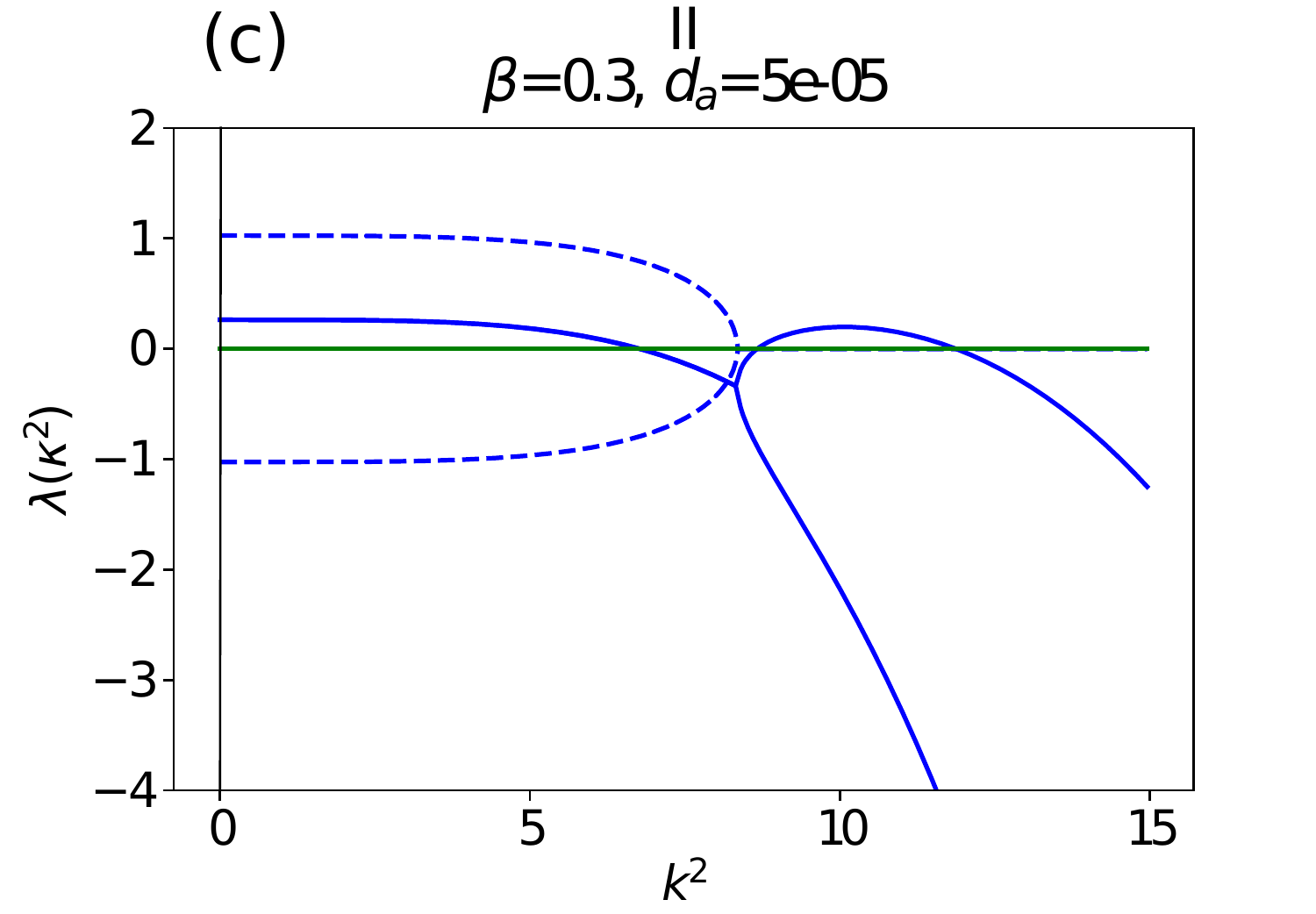} 
	\includegraphics[width=3.2in]{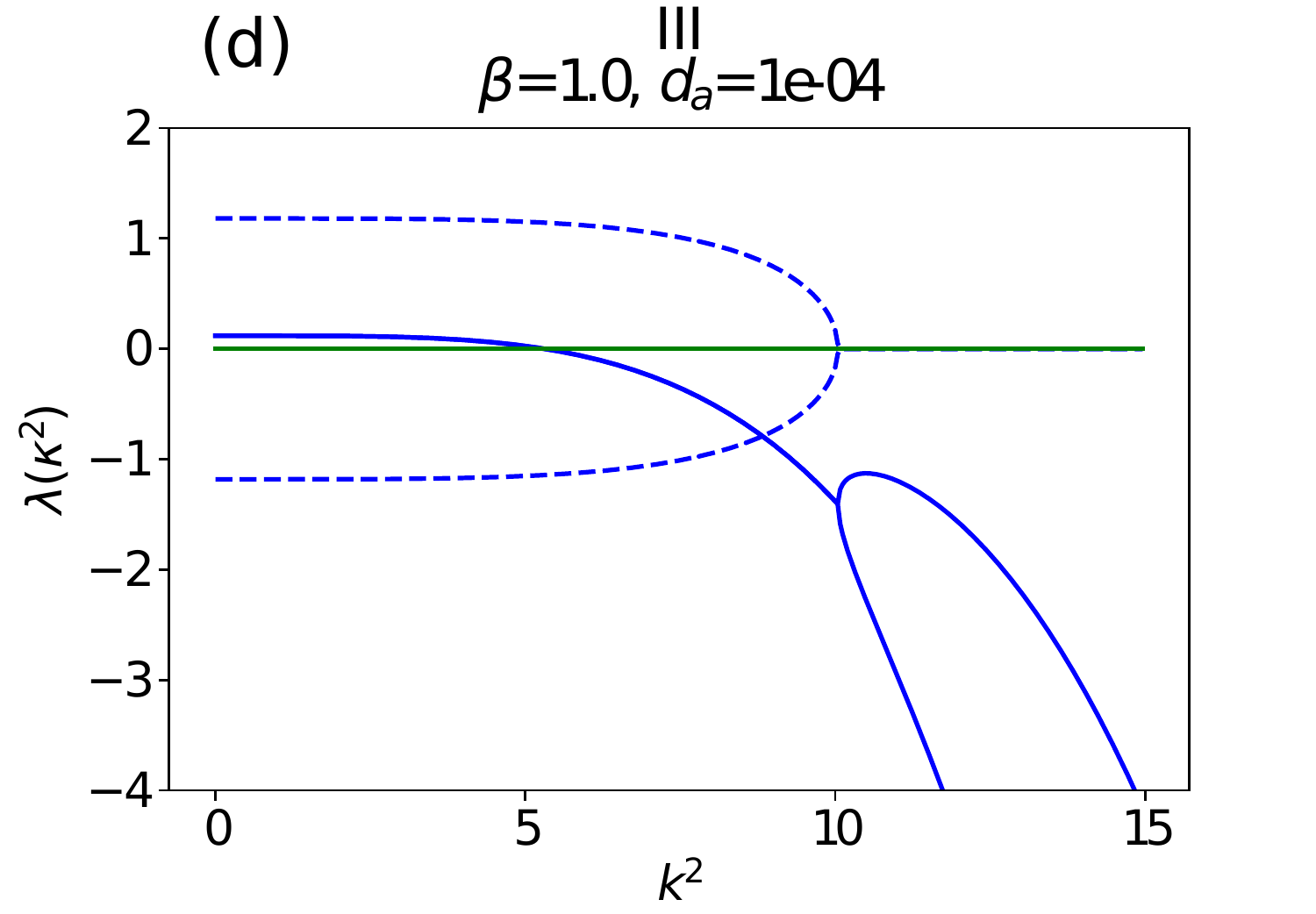} 
	\caption{Samples of dispersion relations for each region depicted in Fig.~\ref{FigB01}. The real parts of the eigenvalues are in solid lines, and the imaginary parts in dashed lines. Panel (a) corresponds to region IV, where the real part is negative for all $\kappa^2$; in panel (b) the real part has two roots, which gives place to a Turing type stationary pattern; in panel (c), Turing and Hopf bifurcations occur for different wave modes: the real part line is positive for wave modes corresponding to non-zero imaginary part, and a Turing instability occurs as in panel (b); in panel (d), a typical Hopf bifurcation feature is exhibited as the real part is positive only for non-zero portions of the imaginary part. The values of $d_a$ and $\beta$ are respectively shown in each panel, and other parameters values are $k_1=5\times10^{-2}, k_2=10^{-2}, k_3=2.0, n=3.0$.}
	\label{FigB02}
\end{figure}

The four kinds of patterns depicted in Fig.~\ref{FigB01} are classified accordingly to the roots of~\eqref{eq:lambda}. A sample of each root-solution type is plotted in Fig \ref{FigB02}, where solid curves are the real part of the eigenvalues and the dashed curves correspond to the imaginary part. The Turing pattern region is labelled by I; as can be seen there, the dispersion relation has a finite positive maximum which gives place to an interval of unstable wave modes. For region II, a Turing-Hopf pattern dispersion relation is depicted, where the imaginary part is non zero for positive portions of the real part, and when the imaginary part colapses to zero, the real part is as in the Turing region. That is, this stability feature gathers two main dynamical ingredients: oscillatory and non oscillatory unstable modes in two finite disconnected subintervals. In region III, a Hopf pattern is characterized by having a dispersion relation in which the real part is only non negative for wave modes where the imaginary part is non zero. Finally, region IV corresponds to no pattern, which is characterized by having a negative real part of the dispersion relation for all wave modes, and hence the homogeneous steady state suffers no symmetry breaking. In other words, the combination of real and imaginary parts of $\lambda(\kappa^2)$ in~\eqref{eqnB06}, provided by~\eqref{eqnB13}, determines each pattern type as well as transitions between regions. For a detailed discussion about this approach, see~\cite{liu}.

This result not only sheds light on the coupling features that diffusion bring about, but also gives further insight on the somite pattern-formation mechanism. That is, region~II in Fig.~\ref{FigB01} corresponds to a dynamic behavior which consists of coordinated transitory oscillations  followed by a stationary state. Such a dynamical response unveils the destabilizing nature of activator's diffusivity; in other words, self-sustained spatially stable oscillations for parameter values in region III become unstable when $\beta$, for a distinguished fixed $d_a$-value, is slowly decreased to land on region~II, where a Turing-type pattern is formed in the end. This transition can be seen in the dispersion relations as in Fig.~\ref{FigB02}(d) that are shaped onto those as in Fig.~\ref{FigB02}(c) as $\beta$ and/or $d_a$ are slowly varied from region III to region II. Observe in both figures that the real part of the eigenvalues start by having positive values for wave modes that correspond to non-zero imaginary parts, see Fig.~\ref{FigB02}(d), to having positive values in two disjointed wave modes intervals, where the imaginary parts are non-zero and null, respectively, as is depicted in Fig.~\ref{FigB02}(c). As a consequence, as a key attribute of the THB, spatially stable limit-cycles are no longer stable to give place to spatially periodic stationary states.

	\begin{figure}[t!]
 		\includegraphics[width=3.2in]{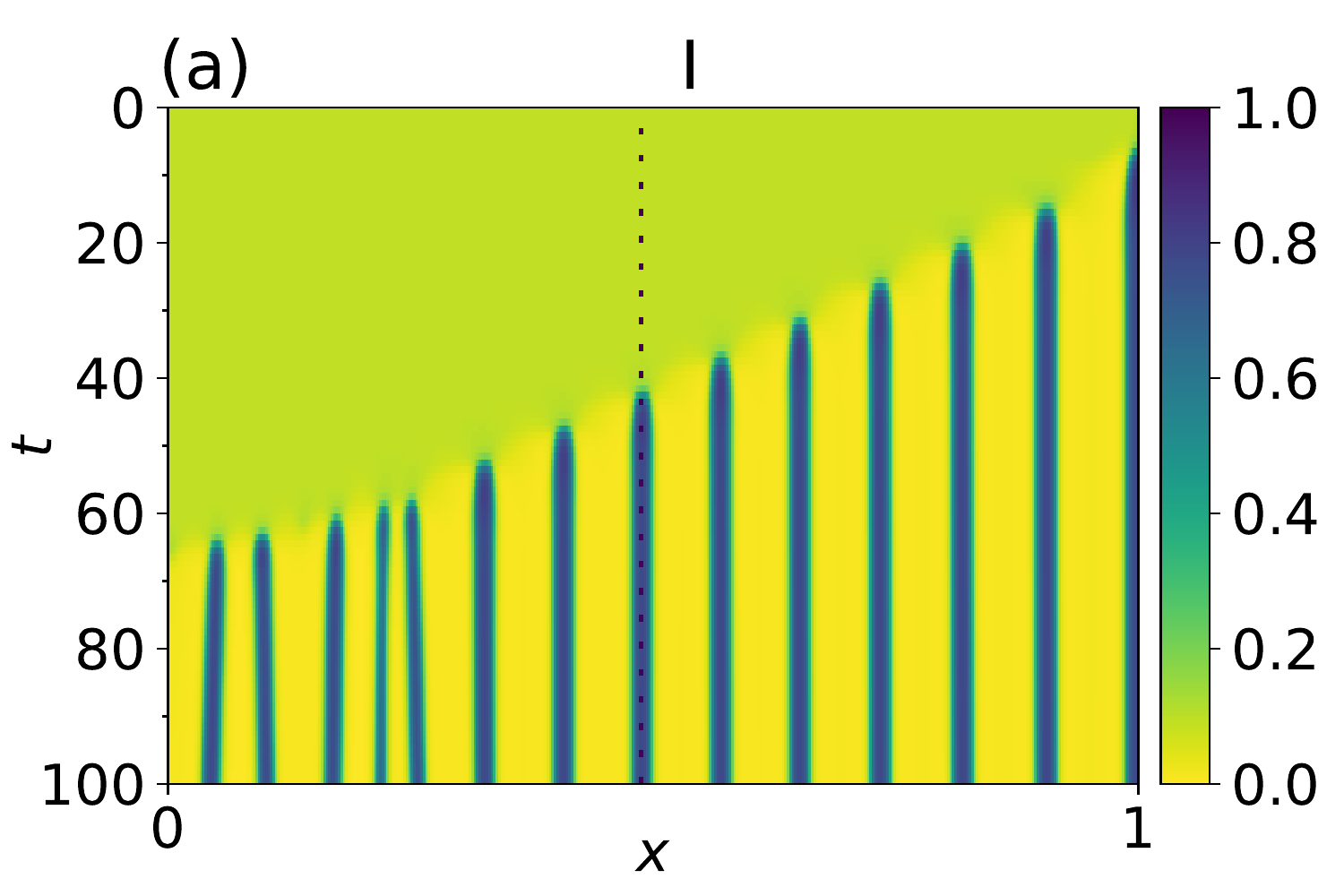}
		\includegraphics[width=3.2in]{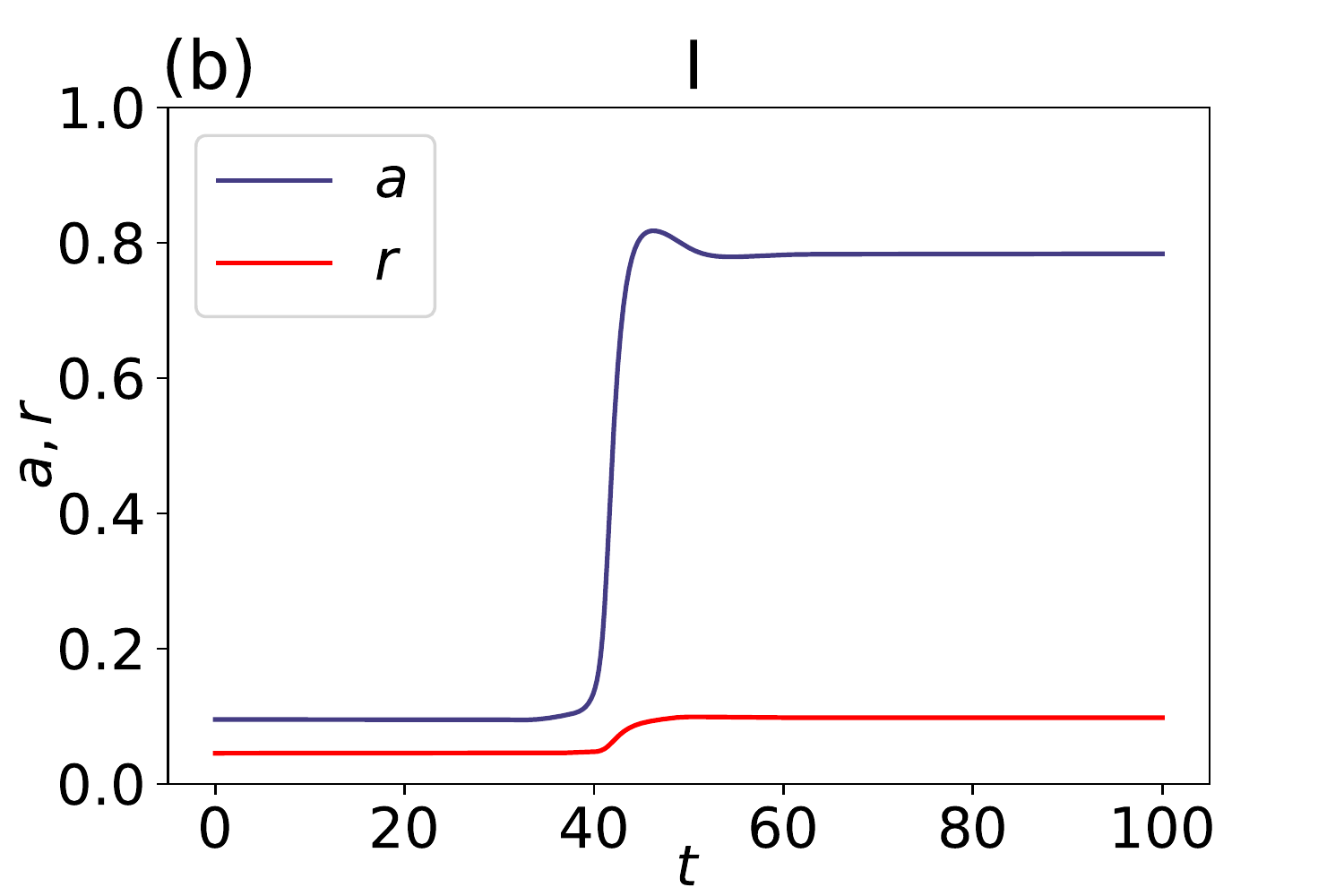}\\
		\includegraphics[width=3.2in]{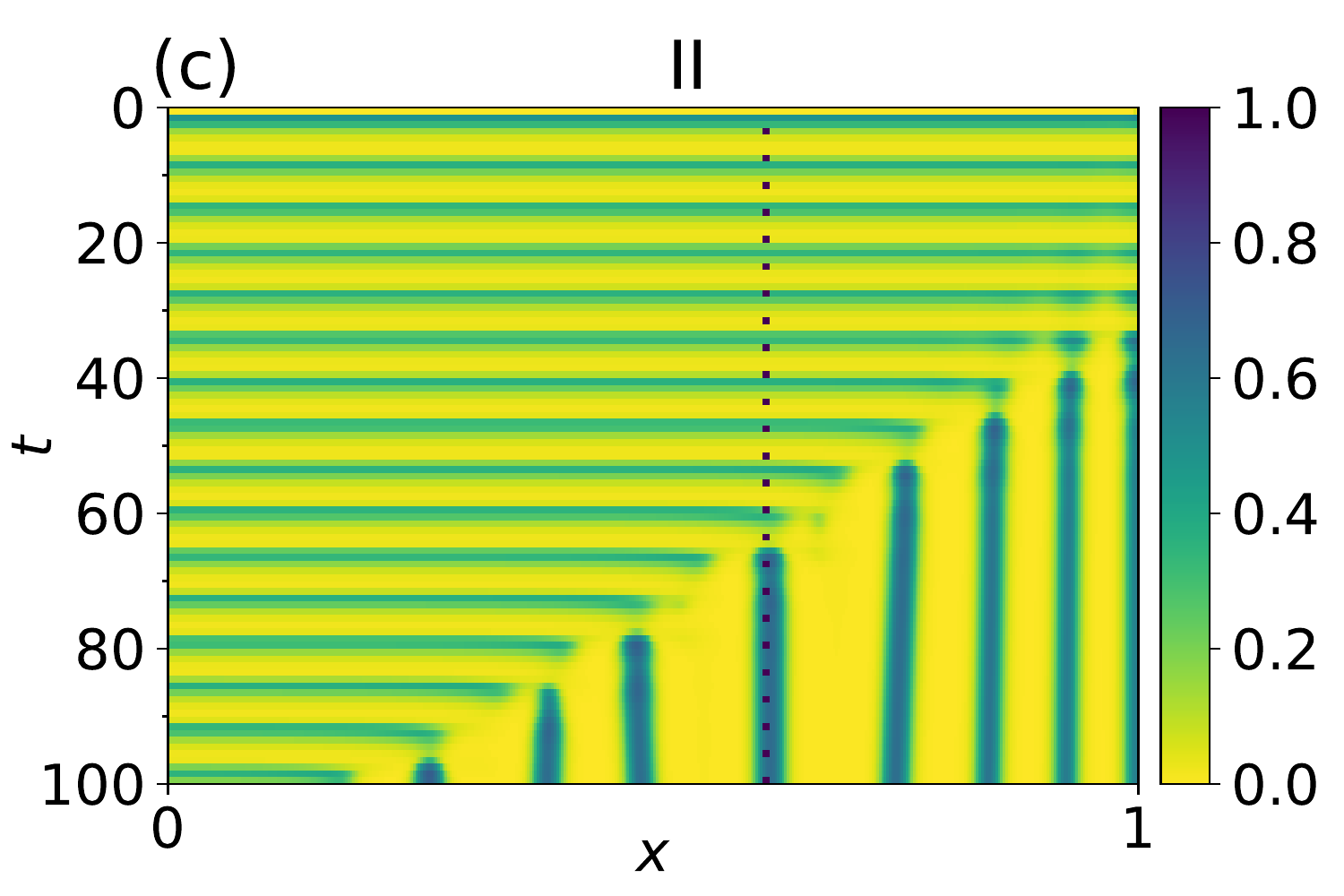}
		\includegraphics[width=3.2in]{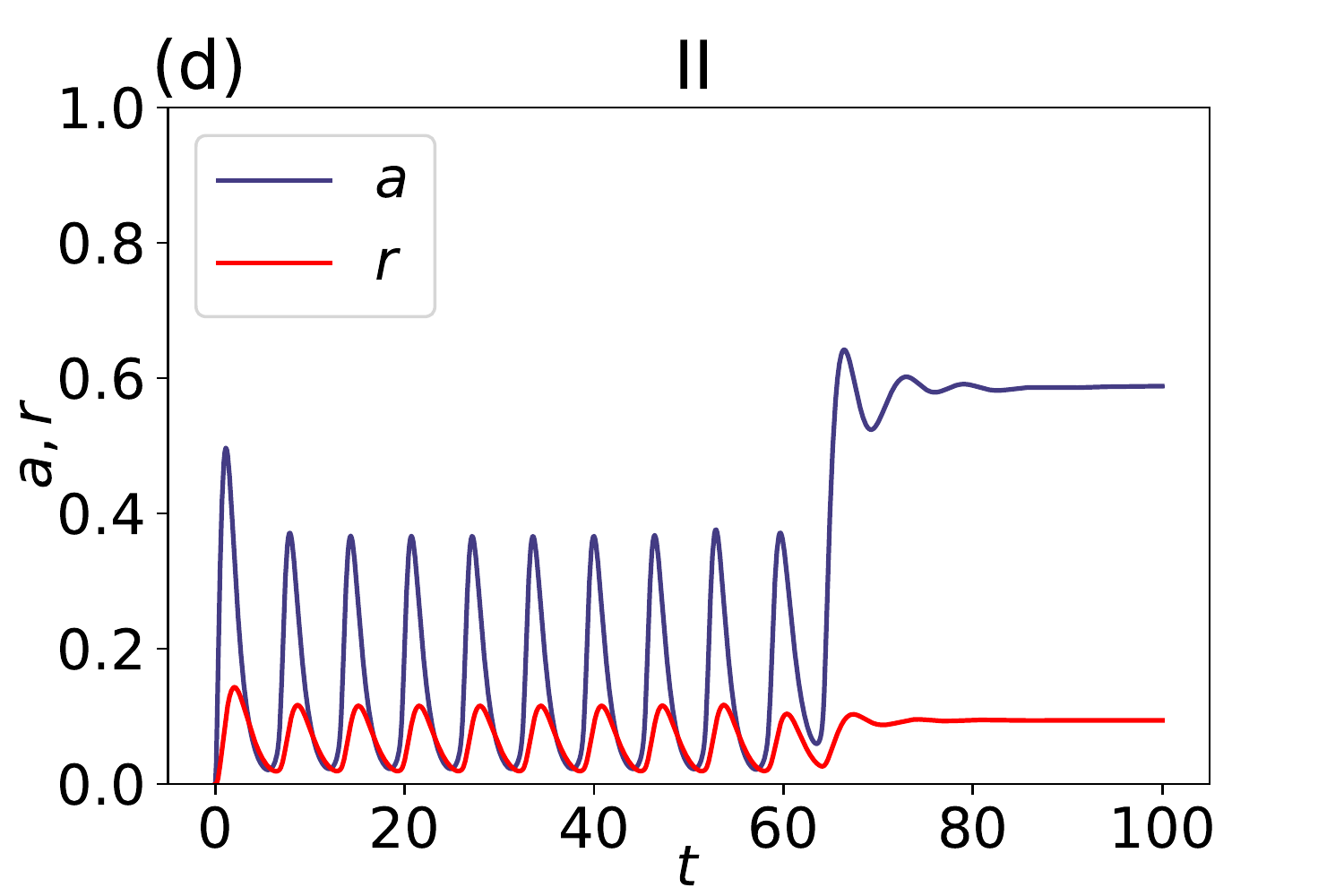}\\
		\caption{Time-step simulations of~\eqref{eqnB01234} with homogeneous Neumann boundary conditions and parameter set values as in region I and II plotted in~Fig.~\ref{FigB01}. An azimuthal view of the spatio-temporal solution shows the pattern formation dynamics for a Turing type (panel a) and a Turing-Hopf type (panel c). Temporal evolution for each case in the left-hand side column of the activator and repressor at $x=0.6125$ and $x=0.4875$ (dashed lines in panels a and c) are shown in panels (b) and (d), respectively. Initial conditions were taken as a perturbation of the steady state accordingly to regions~I~and~II, respectively, of the parameter space in Fig.~\ref{FigB01}. Animations with the same 
			data sets used to plot the heat-maps in this figure can be found in the 
			reservoir \url{https://github.com/JesusPantoja/Reaction-Diffusion_Movies/}.}
		\label{FigB03}
	\end{figure}
	
	\begin{figure}
		\centering
		\includegraphics[width=3.2in]{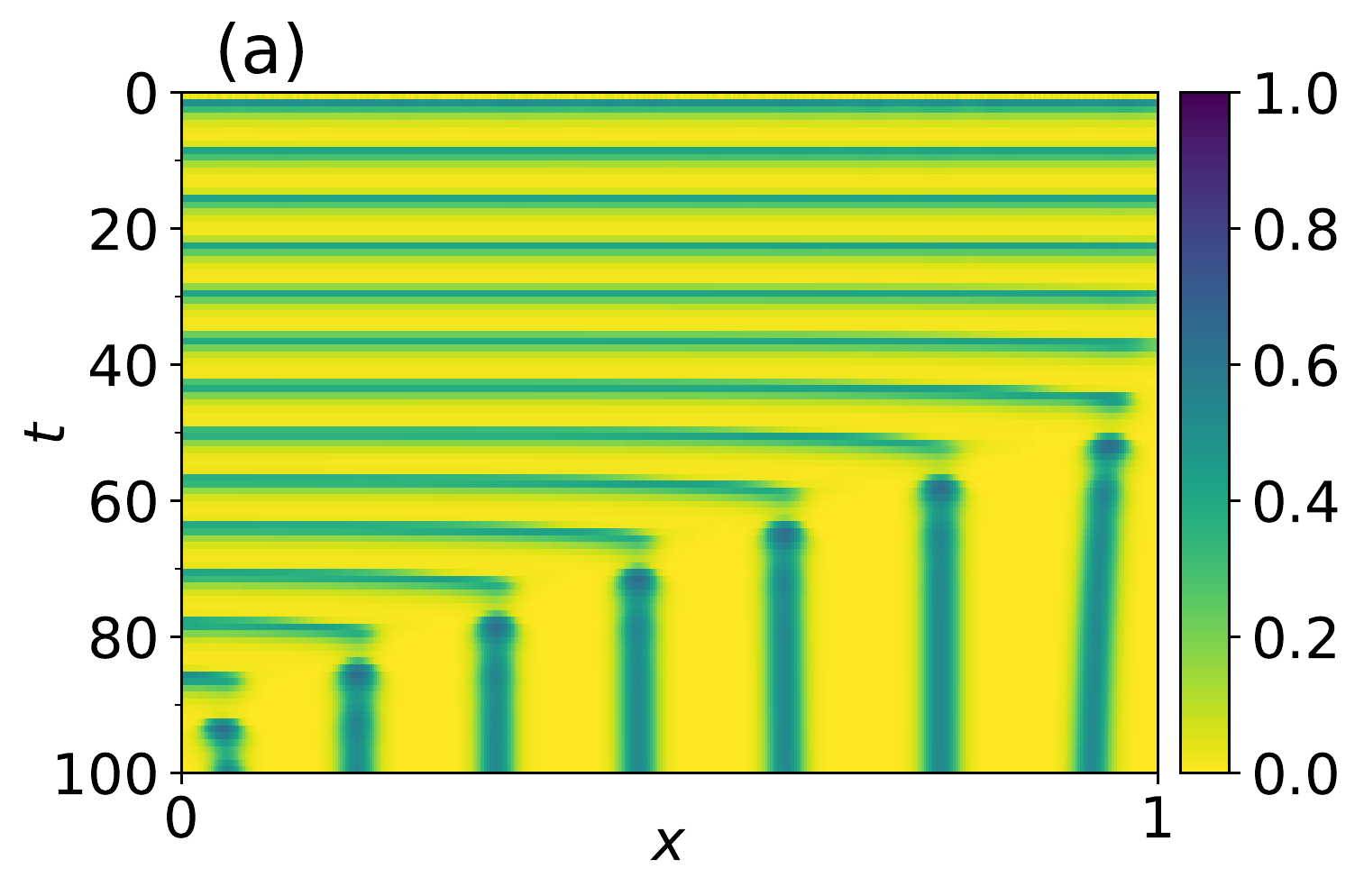}
		\includegraphics[width=3.2in]{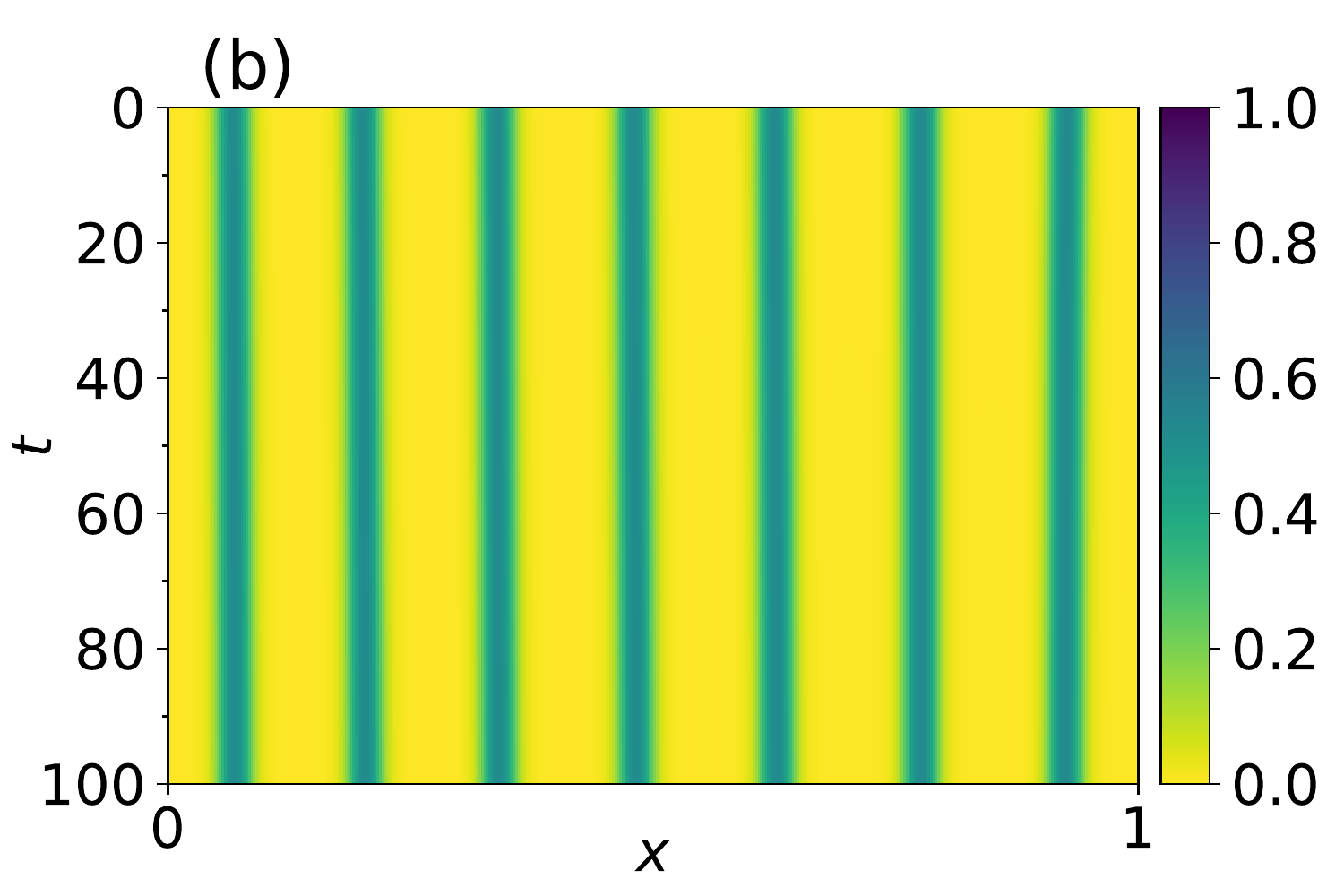}
		\caption{Simulations with parameter $\beta$ as function of $x$ and $t$ as indicated in the text. For (a), the initial conditions are perturbations of the steady state,
		while in (b) the initial conditions are the final state of (a). Animations with the same 
			data sets used to plot the heat-maps in this figure can be found in the 
			reservoir \url{https://github.com/JesusPantoja/Reaction-Diffusion_Movies/}.}
		\label{FigB04}
	\end{figure}

On the other hand, regions I and II are particularly relevant as a stationary pattern is formed, although a key mark lies on transitory dynamical behavior for each scenario. To illustrate this distinguished mark, we perform time-step runs for a setting in both regions, by having a perturbed steady state as an initial condition; see top panels in Fig.~\ref{FigB03}. As can be seen in panel (a), the system is initially in a homogeneous steady state with a slight perturbation. As time goes by, a heterogeneous pattern arises. This is a consequence of the unstable wave modes as is shown in~Fig.~\ref{FigB02}, panel (b), which corresponds to the Turing pattern, region I, in Fig.~\ref{FigB01}. In addition, in panel (b), a time evolution is observed for the activator and repressor states at $x=0.6125$. On the other hand, in an analogous fashion, the transitory dynamics spontaneously oscillate as a consequence of the non zero imaginary part of $\lambda(\kappa^2)$. Such oscillatory dynamics goes on until unstable wave modes allow a stationary pattern to arise. Notice that this mark is clearly observed in panel (d); see bottom panels in Fig.~\ref{FigB03}. In other words, even though both dynamical configurations give place to stationary patterns, the crucial oscillating feature previous to finally get a fixed pattern is added by having a Turing-Hopf mechanism in play.
	
In Fig \ref{FigB04}, we show additional results, where a spatio-temporal dependent parameter $\beta$ is taken into consideration. There, we take $\beta$ as in (\ref{eqbeta}) 
where $v=0.02$ and $v=-0.02$ for the run in panels (a) and (b), respectively. The initial conditions for panel (b) corresponds to the final profile of panel (a). Notice that once the pattern is completely formed, even though $\beta$ varies in a inverse direction, the pattern is not destroyed. This is typical trait of a hysteretical process. In other words, this result indicates that the proposed mechanism is robust since, once the wave front depicted by $\beta$ in (\ref{eqbeta}) prompts the formation of somites, this process cannot be undo.

\clearpage

\bibliographystyle{ieeetr}
\bibliography{ReactionDiffusion}

\end{document}